\documentclass{JHEP3} \usepackage{amssymb} \usepackage{amsfonts} \usepackage{mathrsfs}

\newcommand{\e}{\begin{eqnarray}} \newcommand{\ee}{\end{eqnarray}}

\newcommand{\CN}{{\cal N}} \newcommand{\CL}{{\cal L}}

\def\a{\alpha} \def\b{\beta} \def\d{\delta}
\newcommand{\ep}{\epsilon} \newcommand{\g}{\gamma} \newcommand{\om}{\omega}
\newcommand{\p}{\psi} \newcommand{\s}{\sigma} \def\t{\tau} \newcommand{\vp}{\varepsilon}

\newcommand{\pa}{{a^{\prime}}} \newcommand{\pb}{{b^{\prime}}} \newcommand{\pc}{{c^{\prime}}}
\newcommand{\pd}{{d^{\prime}}} \newcommand{\pe}{{e^{\prime}}} \newcommand{\prf}{{f^{\prime}}}
\newcommand{\pg}{{g^{\prime}}} \newcommand{\pri}{{i^{\prime}}} \newcommand{\pj}{{j^{\prime}}}
\newcommand{\pk}{{k^{\prime}}} \newcommand{\pl}{{l^{\prime}}}

 \newcommand{\pu}{{u^{\prime}}} \newcommand{\pv}{{v^{\prime}}}
\newcommand{\pw}{{w^{\prime}}}  

\newcommand{\gap}{{\alpha^{\prime}}} \newcommand{\gbp}{{\beta^{\prime}}}
\newcommand{\ggp}{{\gamma^{\prime}}}

\newcommand{\DA}{{\dot A}} \newcommand{\DB}{{\dot B}} \newcommand{\DC}{{\dot C}}
\newcommand{\DD}{{\dot D}}

\newcommand{\bi}{\bar{i}}
\newcommand{\bj}{\bar{j}} \newcommand{\bk}{\bar{k}} \newcommand{\bl}{\bar{l}}

\newcommand{\bt}{\bar{t}} \newcommand{\bu}{\bar{u}} \newcommand{\bv}{\bar{v}}
\newcommand{\bw}{\bar{w}}  

\newcommand{\bp}{\bar{\psi}}

\newcommand{\hi}{\hat{i}}
\newcommand{\hj}{\hat{j}} \newcommand{\hk}{\hat{k}} \newcommand{\hl}{\hat{l}}

\newcommand{\tu}{\tilde{u}}

 \title{ Construction of New $D=3, {\cal N}=4$ Quiver Gauge
Theories}

\author{Fa-Min Chen$^{1}$, Yong-Shi Wu$^{2,3}$ \\

${}^1$ Department of Physics, Beijing Jiaotong University, Beijing 100044, China \\
${}^2$ Department of Physics and Center for Field Theory and Particle Physics, \\ Fudan University, Shanghai 200433, China\\
${}^3$ Department of Physics and Astronomy, University of Utah\\ Salt Lake City, UT
84112-0830, USA \\
E-mail: \email{fmchen@bjtu.edu.cn}, \email{wu@physics.utah.edu}}

\abstract{In this paper we propose a special class of 3-algebras, called double-symplectic 3-algebras. We further show that a consistent contraction of the double-symplectic 3-algebra gives a new 3-algebra, called an ${\cal N}=4$ three-algebra, which is then identified as the exact gauged three-algebra in the ${\cal N}=4$ quiver gauge theories.  A systematic construction is proposed for the 3-brackets and fundamental identities used in building up the $\CN=4$ theories, by starting with two superalgebras
whose bosonic parts share at least one simple factor or $U(1)$ factor. This leads to a systematic way of constructing $D=3$, ${\cal N}=4$ quiver theories, of which several examples with new gauge groups are presented in detail. The general ${\cal N}=4$ superconformal Chern-Simons matter theories in terms of ordinary Lie algebras can be also re-derived in our new 3-algebra approach.}


\usepackage{bbm} \usepackage{amsmath} \usepackage{slashed}

\begin{document}

\section{Introduction} \label{Introduction}

In recent years, the construction of extended supersymmetric $(\CN\geq 4)$
Chern-Simons-matter (CSM) gauge theories has received much attention, because they are
candidate low-energy description of multiple M2-branes in M-theory
\cite{Bagger}-\cite{Chen:pku1}. There are essentially two different approaches: (1) The
three-algebra approach \cite{Bagger, Gustavsson, Bagger08:3Alg, ChenWu1, Chen2, ChenWu3,
Bagger10, Chen:pku1}, and (2) The ordinary Lie algebra approach \cite{GaWi, ABJM,
HosomichiJD, Bergshoeff, Hosomichi:2008jb, MFM:Aug09}.

Four years ago, a large class of ${\cal N}=4$ superconformal Chern-Simons-matter (CSM)
theories were constructed by Gaiotto and Witten (GW) \cite{GaWi}, using a method that
enhances an ${\cal N}=1 $ supersymmetry to ${\cal N}=4$. They also proved that the gauge
groups of the GW theory can be classified in terms of classical supergroups. In Ref.
\cite{HosomichiJD}, the GW theory was generalized to an $\CN=4$ quiver gauge theory by adding
additional twisted hyper-multiplets; and the original $\CN=8$ theory \cite{Bagger,Gustavsson}
was reformulated as generalized GW theory with $SO(4)$ gauge group. The $\CN=4$ theory constructed by Hosomichi, Lee, Lee, Lee, and Park in Ref. \cite{HosomichiJD} will be called an $\CN=4$ HLLLP theory. In Ref.
\cite{Bergshoeff}, some examples of the $\CN=4$ theories were derived by taking a conformal
limit of $D=3$ gauged supergravity theories. The progress on the $\CN \geq4$ theories
mentioned above was made by using mainly ordinary Lie algebras.

The authors of the present paper have been able to develop the 3-algebra approach, which was
originally proposed \cite{Bagger, Bagger08:3Alg} for constructing the $D=3$, $\CN=6,8$ CSM
theories, into a unifying approach \cite{ChenWu1, Chen2, ChenWu3, Chen:pku1} that can be used
to construct all known CSM theories with extended supersymmetries $\CN=4,5,6,8$. Our approach
is based on introducing a symplectic form into the underlying 3-algebra, which we called a
symplectic 3-algebra.

We observed that the superspace formulation of the general $\CN=4$ CSM quiver gauge theories
in our paper \cite{ChenWu3} is associated with a special class of symlectic 3-algebras. In
this paper we present a general formulation of this class of symplectic algebras, and show
that its consistent contraction gives a new class of symplectic 3-algebras, which can be used
to construct the general $\CN=4$ theories as well. We then construct some  $\CN=4$
theories with new gauge groups and recover all known $\CN=4$ theories by using superalgebras to construct the
3-brackets and fundamental identities used in building up the $\CN=4$ theories. It is also
demonstrated that the general ${\cal N}=4$ theories \cite{HosomichiJD} in terms of Lie 2-algebras can be
rederived in our 3-algebra approach.

Generators of the 3-algebra in this class are a disjoint union of those of \emph{two}
symplectic sub 3-algebras, whose generators are $T_a$ and $T_\pa$ ($a=1,\cdots,2R$ and
$\pa=1,\cdots, 2S$), respectively. Hence we call it a \emph{double-symplectic 3-algebra}.
(Its contraction will give an \emph{$\CN=4$ three-algebra}.) In this paper we allow the
number of generators in the two sub 3-algebras to be unequal. The double-symplectic 3-algebra
contains \emph{six} independent 3-brackets whose structure constants satisfy \emph{eight}
independent fundamental identities (FIs) and possess certain symmetry property (see Section
\ref{sec2sub}).  The associated matter multiplets, called the un-twisted and twisted
multiplets, take values in the two symplectic sub 3-algebras, i.e. \e\label{3algvalued}
\Phi_A=\Phi^a_AT_a \quad {\rm and}\quad \Phi_\DA=\Phi^\pa_{\DA}T_\pa, \ee
respectively, where $A=1,2$ and $\DA=1,2$ are indices for fundamental representation of the
$SU(2)\times SU(2)$ R-symmetry group.

Then in accordance with the general rules presented in previous papers \cite{Bagger,
ChenWu1}, the general $\CN=4$ theories can be built up by (partially) gauging the symmetry
associated with the double-symplectic 3-algebra \cite{ChenWu3}, if the structure constants
satisfy certain reality conditions and constraint equations. Specifically, we defined
carefully the symmetry transformations that were being gauged, so that after the symmetry
transformations \emph{$\Phi_A$ ($\Phi^\prime_\DA$) still take value in the sub 3-algebra
spanned by $T_a$ ($T_\pa$).} Under this condition, in Ref. \cite{ChenWu3} we used only
structure constants from \emph{four} 3-brackets out of six 3-brackets in constructing the
$\CN=4$ Lagrangian, and used only \emph{four} FIs (see (\ref{FI4})) out of eight to
demonstrate the closure of the gauged symmetry transformations. In this paper we formalize
the gauged symmetry as generated by a new 3-algebra, called the $\CN=4$ 3-algebra, as a
contraction of the original double symplectic 3-algebra as follows: The $\CN=4$ three-algebra
has the the same set of generators    $T_a$ and $T_\pa$, with the following four 3-brackets
remaining unchanged:
 \e &&[T_a,T_b;T_c]=f_{abc}{}^d T_d, \quad
[T_\pa,T_\pb;T_\pc]=f_{\pa\pb\pc}{}^\pd T_\pd, \label{mbarcks0}\\
&&[T_a,T_b;T_\pc]=f_{ab\pc}{}^\pd T_\pd, \quad [T_\pa,T_\pb;T_c]=f_{\pa\pb c}{}^d T_d,
\label{mbrcks00} \ee and the four FIs (see (\ref{FI4})) remaining unchanged as well. But the
structure constants of the remaining two 3-brackets of the $\CN=4$ three-algebra are set to
zero; we call this procedure as contraction of the starting double symplectic 3-algebra. The
contraction turns out to be consistent in the sense that the remanining four FIs are
automatically satisfied due to the vanishing structure constants. In this way, we identify
the \emph{ad hoc} symmetry that was gauged in our previous construction \cite{ChenWu3} of
$\CN=4$ theories as a 3-algebra symmetry generated by the resulting $\CN=4$ 3-algebra after
the consistent contraction.

We briefly describe how to construct the four sets of 3-brackets (\ref{mbarcks0}) and
(\ref{mbrcks00}) and the four sets of FIs (\ref{FI4}) in terms of two superalgebras. A
sympelctic 3-algebra can be realized in terms of a superalgebra \cite{ChenWu3}. It is natural
to introduce \emph{two} superalgebras to realize the \emph{two} symplectic sub 3-algebras. We
first introduce a superalgebra $G$ (see (\ref{slie2})) to realize one of the two sympelctic
three-subalgebras: \begin{equation} T_a\doteq Q_a, \quad [T_a, T_b; T_c]\doteq [\{Q_a, Q_b\},
Q_c]. \end{equation} where $Q_a$ are fermionic generators of $G$. We have identified the
generators of the sub 3-algebra with the fermionic generators of $G$ and constructed the
3-bracket in terms of a double graded commutator on $G$. In this realization, one can
formulate the fundamental identity (FI) of the sub symplectic 3-algebra as the $M_uM_vQ_e$
Jacobi identity of the superalgebra $G$ ($M_u$ are the bosonic generators of $G$), and the
constraint equation $f_{(abc)d}=0$ required for enhancing the $\CN=1$ supersymmetry to
$\CN=4$ is equivalent to the $Q_aQ_bQ_c$ Jacobi identity, and the structure constants
$f_{abc}{}^d$ satisfy the correct symmetry and reality conditions as well. Similarly, we
introduce another superalgebra $G^\prime$ to realize another sub symplectic 3-algebra:
\begin{equation} T_\pa\doteq Q_\pa, \quad [T_\pa, T_\pb; T_\pc]\doteq [\{Q_\pa, Q_\pb\},
Q_\pc]. \end{equation}

Finally, it is natural to construct the 3-brackets (\ref{mbrcks00}) in terms of double graded
commutators defined on the superalgebras $G$ \emph{and} $G^\prime$, for instance,
\begin{equation} [T_a,T_b;T_\pc]\doteq [\{Q_a, Q_b\}, Q_\pc]. \end{equation} In order that
there are non-trivial interactions between the twisted and un-twisted multiplets, we must
require that \e [T_a,T_b;T_\pc]\neq0,\quad [T_\pa,T_\pb;T_c]\neq0. \ee As we will prove, a
sufficient condition for $[\{Q_a, Q_b\}, Q_\pc]\neq 0$ and $[\{Q_\pa, Q_\pb\}, Q_c]\neq 0$ is
that the bosonic parts of $G$ and $G^\prime$ share at least one simple factor or $U(1)$
factor. Two fundamental identities obeyed by the two structure constants of the two
3-brackets in (\ref{mbrcks00}) (see the second and third equation of (\ref{FI4})), follow
from the $M_uM_vQ_\pe$ and $M_\pu M_\pv Q_e$ Jacobi identities, respectively. Here $M_\pu$
are bosonic generators of $G^\prime$. In this realization, the Lie algebra of the gauge group
of the $\CN=4$ quiver theory is the bosonic subalgebras of $G$ \emph{and} $G^\prime$, and the
corresponding matter representation is determined by the fermionic generators $Q_a$ and
$Q_\pb$. For example, we may choose $G=OSp(2|2N)$ and $G^\prime=U(M|1)$, whose bosonic parts
$Sp(2N)\times U(1)$ and $U(1)\times U(M)$ share a common factor $SO(2)\cong U(1)$. Then the
resulting gauge group is $Sp(2N)\times U(1)\times U(M)$, and the un-twisted multiplet is in
the bi-fundamental representation of $Sp(2N)\times U(1)$, while the twisted multiplet is in
the bi-fundamental representation of $U(1)\times U(M)$. This example is one of our new
$\CN=4$ theories. Some new $\CN=4$ theories may contain free parameters; we construct two
infinite classes of new theories of this kind in Sec. \ref{sec4un} and Sec. \ref{secospu}.

Since if we use superalgebras
to construct the 3-algebras, the general $\CN=4$ theories in terms of 3-algebras are equivalent to the
$\CN=4$ HLLLP theories based on conventional Lie 2-algebras, the $\CN=4$ theories with new gauge groups (see Sec. \ref{secNewEx} and \ref{Allex}) can be also
derived directly by taking advantage of the classification of gauge groups of the general
$\CN=4$ HLLLP theories \cite{HosomichiJD}. The essentially new things here are only the gauge groups, which were not explicitly discussed in the literature. It is in this sense that we
construct new $\CN=4$ theories.

The superalgebras $G$ and $G^\prime$, whose bosonic parts share at least a simple factor or a
$U(1)$ factor, are interested in their own right. In a coming paper \cite{ChenWu5}, we will
propose a procedure to ``fuse" certain $G$ and $G^\prime$ into a single closed superalgebra.

This paper is organized as follows. In section \ref{sec2sub}, we introduce the
double-symplectic 3-algebra and the $\CN=4$ three-algebra, and review the  $\CN=4$ theories
based on the 3-algebras in section \ref{secN4}. In section \ref{srt0}, we construct the four
sets of 3-brackets (\ref{mbarcks0}) and (\ref{mbrcks00}) and the four sets of FIs (\ref{FI4})
in terms of two superalgebras, (We also comment the rest two 3-brackets and four FIs which do
not play roles in constructing the $\CN=4$ theory.) and the subsequent section \ref{secNewEx}
is devoted to present the construction of a number of new classes of $\CN=4$ theories. In
section \ref{secGN4}, the general ${\cal N}=4$ CSM theories in terms of ordinary Lie algebras
are derived from their 3-algebra counterpart. In section \ref{Allex}, we recover all known
examples of $\CN=4$ theories, and produce more new examples. We end in section
\ref{conclusions} with conclusions and discussions. We summarize our conventions in Appendix
\ref{Identities}. The commutation relations of some superalgebras used to construct
symplectic sub 3-algebras are given in Appendix \ref{superalgebras}.

\section{Double-Symplectic 3-Algebra and $\CN=4$ Three-Algebra}\label{sec2sub}

In this section, we introduce the \emph{two} classes of 3-algebras mentioned in Section
\ref{Introduction}. Both of them can be used to construct the general $\CN=4$ quiver gauge
theory. We first introduce the \emph{double-symplectic 3-algebra} whose generators are the
\emph{disjoint union} of those of two sympelectic sub 3-algebras. We then introduce the
\emph{$\CN=4$ three-algebra} as a consistent contraction of the double-symplectic 3-algebra.
The self-consistency of the $\CN=4$ three-algebras is revealed explicitly.

We denote the generators of the two sub 3-algebra as $T_a$ and $T_\pa$, respectively, where
$a=1,\cdots,2R$ and $\pa=1,\cdots, 2S$. 
To define two
symplectic sub 3-algebras, one must introduce two invariant anti-symmetric tensors
\e\label{forms0} \omega_{ab}=\omega(T_a, T_b)\quad {\rm and}\quad
\omega_{\pa\pb}=\omega(T_\pa, T_\pb) \ee into the two sub 3-algebras, respectively. We denote
their inverses as $\omega^{bc}$ and $\omega^{\pb\pc}$, satisfying
$\omega_{ab}\omega^{bc}=\delta^c_a$ and $\omega_{\pa\pb}\omega^{\pb\pc}=\delta^\pc_\pa$. 
$\omega^{bc}$ and $\omega^{\pb\pc}$ are required to be invariant under the transformations
(\ref{glbtran1u}) and (\ref{glbtran1p}), respectively. The antisymmetric tensors $\omega$
will be used to lower or raise the indices. Also, we require that the unprimed and primed
generators to be symplectic orthogonal, in the sense that \e\label{ortho} \omega(T_a,
T_\pb)=\omega(T_\pb, T_a)=0. \ee 


Since $T_a$ \emph{and} $T_\pa$ form a complete basis, the general 3-bracket on this 3-algebra
can be defined as \begin{eqnarray}\label{Symp3Bracket} [T_I,T_J;T_K]&=&f_{IJK }{}^dT_{d}+
f_{IJK }{}^\pd T_{\pd}\\ &\equiv& g_{IJK}{}^LT_L,\nonumber \end{eqnarray} where $T_I, T_J$
and $T_K$ are arbitrary three generators selected from the 3-algebra; each of them can be a
primed \emph{or} an unprimed generator, for instance, \e\label{compact} T_I=(T_a\!&{\rm
or}\!& T_\pa). \ee The basic property of the 3-bracket of symplectic 3-algebra is that it is
invariant if we switch the first two generators \cite{Chen2}, i.e. \e\label{symin2}
[T_I,T_J;T_K]=[T_J,T_I;T_K]. \ee And we assume that the structure constants satisfy the
symmetry condition \e\label{bigsym} \omega({[T_I,T_J;T_K], T_L})=\omega({[T_K,T_L;T_I],
T_J}). \ee The generators are required to satisfy the fundamental identity:
\begin{equation}\label{FI} [T_I,T_J; [T_M,T_N;T_K]]=[[T_I,T_J;T_M],T_N;
T_K]+[T_M,[T_I,T_J;T_N]; T_K]+[T_M,T_N; [T_I,T_J;T_K]]. \end{equation} The FI plays an
analogous role as the Jacobi identity of an ordinary Lie algebra. Substituting
(\ref{Symp3Bracket}) into (\ref{FI}), we obtain the FI satisfied by the structure constants:
\begin{equation}\label{FFI} g_{MNK}{}^Og_{IJO}{}^{L}=g_{IJM}{}^Og_{ONK}{}^{L}
+g_{IJN}{}^Og_{MOK}{}^{L}+g_{IJK}{}^Og_{MNO}{}^{L}. \end{equation}

We call the 3-algebra defined by Eqs. (\ref{forms0}), (\ref{ortho}), (\ref{Symp3Bracket}),
(\ref{symin2}), (\ref{bigsym}) and (\ref{FI}) a \emph{double-symplectic 3-algebra}.

Taking account of (\ref{compact}) and (\ref{symin2}), we see that Eq. (\ref{Symp3Bracket})
represents \emph{six} independent 3-brackets. Since the two sets of generators $T_a$ and
$T_\pa$ span the two sub 3-algebras respectively, the 3-bracket of three unprimed (primed)
generators must be a linear combination of unprimed (primed) generators, i.e.
\begin{equation}\label{own3brcks} [T_a,T_b;T_c]=f_{abc}{}^d T_d\quad{\rm and} \quad
[T_\pa,T_\pb;T_\pc]=f_{\pa\pb\pc}{}^\pd T_\pd. \end{equation} Comparing (\ref{own3brcks})
with the general definition of 3-bracket (\ref{Symp3Bracket}), we notice that
\e\label{vanish} f_{abc}{}^\pd=f_{\pa\pb\pc}{}^d=0. \ee For the rest four 3-brackets, every
one contains one primed (unprimed) generator and two unprimed (primed) generators. However,
taking account of (\ref{ortho}), (\ref{symin2}), (\ref{bigsym}) and (\ref{vanish}), we find
that the four structure constants carrying three unprimed (primed) indices and one primed
(umprimed) vanish, i.e. \e\label{unused0} f_{ab\pc}{}^d=f_{\pa\pb c}{}^\pd=f_{a\pb
c}{}^d=f_{b\pa\pc}{}^\pd=0. \ee  So these four 3-brackets are given by
\e\label{mbrcks} &&[T_a,T_b;T_\pc]=f_{ab\pc}{}^\pd T_\pd,\quad [T_\pa,T_\pb;T_c]=f_{\pa\pb
c}{}^d T_d,\\ &&[T_a,T_\pb;T_c]=f_{a\pb c}{}^\pd T_\pd,\quad [T_\pa,T_b;T_\pc]=f_{ b
\pa\pc}{}^dT_d.\label{mbrcks3}\ee Eqs (\ref{vanish}), (\ref{unused0}) and (\ref{bigsym})
imply that \e\label{bigsym1} f_{IJKL}=f_{KLIJ}, \ee where $f_{abcd}=\omega_{de}f_{abc}{}^e$
and $f_{ab\pc\pd}=\omega_{\pd\pe}f_{abc}{}^\pe$.

Taking account of (\ref{symin2}) and
(\ref{bigsym}), Eq. (\ref{FFI}) is decomposed into \emph{eight} independent FIs. The four
subsets of FIs which do not involve $f_{a\pb c\pd}$ are given by
\e
\label{FI4}&&f_{abe}{}^gf_{gfcd}+
f_{abf}{}^gf_{egcd}-f_{efd}{}^gf_{abcg}-f_{efc}{}^gf_{abdg}=0 ,\nonumber\\
&&f_{abe}{}^gf_{gf\pc\pd}+ f_{abf}{}^gf_{eg\pc\pd}-f_{ef\pd}{}^\pg
f_{ab\pc\pg}-f_{ef\pc}{}^\pg f_{ab\pd\pg}=0 ,\\ &&f_{\pa\pb e}{}^gf_{gf\pc\pd}+ f_{\pa\pb
f}{}^gf_{eg\pc\pd}-f_{ef\pd}{}^\pg f_{\pa\pb\pc\pg}-f_{ef\pc}{}^\pg
f_{\pa\pb\pd\pg}=0,\nonumber\\ &&f_{\pa\pb \pe}{}^\pg f_{\pg\prf\pc\pd}+ f_{\pa\pb
\prf}{}^\pg f_{\pe\pg\pc\pd}-f_{\pe\prf\pd}{}^\pg f_{\pa\pb\pc\pg}-f_{\pe\prf\pc}{}^\pg
f_{\pa\pb\pd\pg}=0. \nonumber\ee
The other four subsets of FIs involving $f_{a\pb c\pd}$ are given by
\e \label{FI5}&&f_{a\pc b}{}^\pd f_{ef\pd\pg}= f_{efa}{}^df_{d\pc
b\pg}+f_{ef\pc}{}^\pd f_{a\pd b\pg}+f_{efb}{}^df_{a\pc d\pg} ,\nonumber\\ &&f_{a\pc b}{}^\pd
f_{e\prf g\pd}= f_{e\prf a}{}^\pd f_{\pd\pc bg}+f_{e\prf}{}^d{}_\pc f_{adbg}+f_{e\prf
b}{}^\pd f_{a\pc g\pd} ,\\ &&f_{a\pc b}{}^\pd f_{\pe\prf\pd\pg}= f_{\pe\prf a}{}^df_{d\pc
b\pg}+f_{\pe\prf\pc}{}^\pd f_{a\pd b\pg}+f_{\pe\prf b}{}^df_{a\pc d\pg} ,\nonumber\\
&&f_{a\pc}{}^d{}_\pg f_{e\prf d\pb}= f_{e\prf a}{}^\pd f_{\pd\pc\pg\pb}+f_{e\prf}{}^d{}_\pc
f_{ad\pg\pb}+f_{e\prf}{}^d{}_\pg f_{a\pc d\pb}. \nonumber\ee

In the previous construction of the $\CN=4$ theories \cite{ChenWu3}, we assumed that the
variation of a 3-algebra valued superfield $\Phi$ takes the form
 \e\label{glbtran1}
\delta_{\tilde\Lambda}\Phi=\Lambda^{ab}[T_a,T_b;\Phi]+\Lambda^{\pa\pb}[T_\pa,T_\pb;\Phi], \ee
where $\Phi$ can be either an untwisted superfield $\Phi=\Phi^a_AT_a$ \emph{or} a twisted
superfield $\Phi=\Phi^\pa_\DA T_\pa$. The infinitesimal parameters $\Lambda^{ab}$ and
$\Lambda^{\pa\pb}$ are independent of superspace coordinates. In terms of components, Eq.
(\ref{glbtran1}) can be written as \e
\delta_{\tilde\Lambda}\Phi^d_A&=&\Lambda^{ab}f_{abc}{}^d\Phi^c_A+ \Lambda^{\pa\pb}f_{\pa\pb
c}{}^d\Phi^c_A\label{glbtran1u}, \\
\delta_{\tilde\Lambda}\Phi^\pd_\DA&=&\Lambda^{ab}f_{ab\pc}{}^\pd\Phi^\pc_A+
\Lambda^{\pa\pb}f_{\pa\pb\pc}{}^\pd\Phi^\pc_\DA.\label{glbtran1p} \ee We also assumed that
the action is invariant under the symmetry transformation (\ref{glbtran1}); and the symmetry
will be gauged later \cite{ChenWu3}.

Eq. (\ref{glbtran1}) is obviously not the most general
possibility,  since one may add another term
like \e\label{hete} \Lambda^{a\pb}[T_a,T_\pb;\Phi]\equiv \delta_{\tilde\Lambda_3}\Phi \ee to
the right hand side of (\ref{glbtran1}). However, substituting $\Phi=\Phi^a_AT_a$ into
(\ref{hete}) and using the first equation of (\ref{mbrcks3}), we obtain \e
\Lambda^{a\pb}[T_a,T_\pb;\Phi^c_AT_c]=\Lambda^{a\pb}f_{a\pb c}{}^{\pd} T_{\pd} \Phi^c_A. \ee
The right hand side indicates that $\delta_{\tilde\Lambda_3}\Phi_A$ does \emph{not} take
value in the sub 3-algebra spanned by the unprimed generators anymore, i.e.
$\delta_{\tilde\Lambda_3}\Phi_A\neq (\d_{\tilde\Lambda_3}\Phi)^c_AT_c$, which conflicts with
our basic assumption that $\Phi_A=\Phi_A^aT_a$. We therefore must exclude this term by
setting either $\Lambda^{a\pb}=0$ \emph{or} $f_{a\pb c}{}^\pd=0$. \begin{itemize} \item If we
set $\Lambda^{a\pb}=0$, we will gauge the symmetry defined by the symmetry
    transformation (\ref{glbtran1}), which
is only \emph{part} of the full symmetry generated by the double-symplectic 3-algebra. \item
If we set $f_{a\pb c}{}^\pd=0$, using (\ref{bigsym}) and (\ref{symin2}), we obtain
    $f_{a\pb\pd}{}^c=0$, implying that $\delta_{\tilde\Lambda_3}\Phi_\DA=0$.
The new 3-algebra, obtained from the double-symplectic 3-algebra by setting $f_{a\pb
c}{}^{\pd}=f_{b\pa\pd}{}^{c}=0$ while keeping the rest structure constants unchanged, will be
called \emph{an $\CN=4$ three-algebra}. It is in this sense that we obtain the $\CN=4$
three-algebra from the double-symplectic 3-algebra by a \emph{contraction}.  This contraction
is consistent, since by setting $f_{a\pb c}{}^{\pd}=f_{b\pa\pd}{}^{c}=0$, the other four
subsets of FIs (\ref{FI5}) are automatically satisfied. The only difference between these two
3-algebras is that in the $\CN=4$ three algebra, $f_{a\pb c}{}^{\pd}$ and $f_{b\pa\pd}{}^{c}$
strictly vanish, while in the double-symplectic 3-algebra, generally speaking, they do not
vanish.
 \end{itemize}

Having defined the symmetry transformations (\ref{glbtran1}) and (\ref{hete}), we now want to
examine the invariance of $\omega_{ab}$ and $\omega_{\pa\pb}$ under these transformations.
Since $T_a$ form a complete basis of the symplectic sub 3-algebra, the antisymmetric tensor
$\omega_{cd}$ must be invariant under the transformation: \e\label{symin20}
\delta_{\tilde\Lambda_1}\omega_{cd}=\Lambda^{ab}(f_{abc}{}^e\omega_{ed}+f_{abd}{}^e\omega_{ce})=0,
\ee In order that $\omega_{cd}$ is also invariant under the transformation (\ref{glbtran1u}),
one must require that \e\label{symin21}
\delta_{\tilde\Lambda_2}\omega_{cd}=\Lambda^{\pa\pb}(f_{\pa\pb c}{}^e\omega_{ed}+f_{\pa\pb
d}{}^e\omega_{ce})=0. \ee Eqs (\ref{symin20}) and (\ref{symin21}) are equivalent to
$f_{abcd}=f_{abdc}$ and $f_{\pa\pb cd}=f_{\pa\pb dc}$, respectively. Similarly, we obtain
$f_{\pa\pb\pc\pd}=f_{\pa\pb\pd\pc}$ and $f_{ab\pc\pd}=f_{ab\pd\pc}$ by requiring the
transformations to preserve $\omega_{\pc\pd}$. These equations are consistent with Eq.
(\ref{bigsym}). In the case of the double symplectic 3-algebra, using (\ref{ortho}), it is
not difficult to prove that \e\label{symin22}
\delta_{\tilde\Lambda_3}\omega_{cd}=\delta_{\tilde\Lambda_3}\omega_{\pc\pd}=0, \ee where
$\delta_{\tilde\Lambda_3}$ is defined by Eq. (\ref{hete}). Eqs. (\ref{symin20}) $\sim$
(\ref{symin22}) indicate that $\omega_{cd}$ ($\omega_{\pc\pd}$) is not only an invariant
tensor on the umprimed (primed) sub 3-algebra, but also an invariant tensor on the
double-symplectic 3-algebra. In the case of the $\CN=4$ three-algebra, Eqs. (\ref{symin22})
are satisfied automatically due to the fact that $f_{a\pb c}{}^{\pd}$ and $f_{b\pa\pd}{}^{c}$
strictly vanish.



By substituting $\Phi=\Phi^a_AT_a$ and $\Phi=\Phi^\pa_\DA T_\pa$ into (\ref{glbtran1}),
respectively, we see that the symmetry transformations involve only the four subsets of
structure constants \footnote{Because of the symmetry condition $f_{ab\pc\pd}=f_{\pc\pd ab}$
(see (\ref{symfs})), there are only three independent subsets of structure constants.}:
 \e\label{4strc} f_{abc}{}^d,\quad f_{\pa\pb\pc}{}^\pd,\quad
f_{ab\pc}{}^\pd\quad {\rm and}\quad f_{\pa\pb c}{}^d. \ee Therefore, only four subsets of
structure constants appear in the action and the supersymmetry transformations (see
(\ref{LN4}) and (\ref{SUSY4})) \cite{ChenWu3}.  In summary, these structure constants enjoy
the symmetry properties (see (\ref{symin2}) and (\ref{bigsym})) \begin{eqnarray}\label{symfs}
&&f_{abcd}=f_{bacd}=f_{badc}=f_{cdab},\nonumber\\
&&f_{ab\pc\pd}=f_{ba\pc\pd}=f_{ba\pd\pc}=f_{\pc\pd ab},\\
&&f_{\pa\pb\pc\pd}=f_{\pb\pa\pc\pd}=f_{\pb\pa\pd\pc}=f_{\pc\pd\pa\pb},\nonumber
\end{eqnarray} and satisfy the reality conditions \cite{ChenWu3}
\begin{equation}\label{rltcndtn}f^{*a}{}_b{}^c{}_d=f^{b}{}_a{}^d{}_c, \quad
f^{*\pa}{}_\pb{}^c{}_d=f^{\pb}{}_\pa{}^d{}_c,\quad
f^{*\pa}{}_\pb{}^\pc{}_\pd=f^{\pb}{}_\pa{}^\pd{}_\pc, \end{equation} for guaranteeing the
positivity of the theory. In addition, in order to close the $\CN=4$ super Poincare algebra,
we need to impose the linear constraints on $f_{abcd}$ and $f_{\pa\pb\pc\pd}$
\begin{equation}\label{Constr3} f_{(abc)d}=0\quad {\rm and} \quad f_{(\pa\pb\pc)\pd}=0.
\end{equation}

Since we use the structure constants (\ref{4strc}) to construct the $\CN=4$ theory, they must
be invariant under the symmetry transformation (\ref{glbtran1}), i.e. \e\label{FI6}
\d_{\tilde\Lambda}f_{abcd}=\d_{\tilde\Lambda}f_{ab\pc\pd}=\d_{\tilde\Lambda}
f_{\pa\pb\pc\pd}=0, \ee A short calculation shows that the \emph{four} subsets of FIs
(\ref{FI4}) are equivalent to Eqs. (\ref{FI6}). In particular, Eqs. (\ref{FI6}) do \emph{not}
involve the other four subsets of FIs (\ref{FI5}) at all.

Thus, to construct the $\CN=4$ gauge theory, we only need the \emph{four} subsets of
3-brackets (\ref{own3brcks}) and (\ref{mbrcks}), together with the \emph{four} subsets of FIs
(\ref{FI4}) associated with them. In other words, we need only to gauge the symmetry
generated by \emph{four} sets of 3-brackets (\ref{own3brcks}) and (\ref{mbrcks}) and the
\emph{four} sets of FIs (\ref{FI4}). Later we will have chance to comment on the two other
subsets of 3-brackets (\ref{mbrcks3})  which do not appear in the action and supersymmetry
transformations, and the involved four sets of FIs (\ref{FI5}). (See the last paragraph of
Section \ref{srt0}).



\section{Construction of $\CN=4$ Theories Based on 3-Algebras }\label{secN4}

In this section, we will present the $\CN=4$ theories constructed by gauging part of symmetry
generated by the double-symplectic 3-algebra, or from another point of view, constructed in
terms of the $\CN=4$ three-algebras following the procedure proposed in \cite{ChenWu3}. The
un-twisted multiplets $(Z^a_A, \p^a_\DA)$, valued in the sub 3-algebra, satisfy the reality
conditions \begin{equation} \bar Z^A_a=\omega_{ab}\ep^{AB}Z^b_B,\quad \bar
\p^\DA_a=\omega_{ab}\ep^{\DA\DB}Z^b_\DB, \end{equation} where $A,\DA=1,2$ transform in the
two-dimensional representation of the $SU(2)\times SU(2)$ R-symmetry group. And the twisted
multiplets $(Z^\pa_\DA, \p^\pa_A)$ satisfy similar reality conditions. The gauge fields and
the covariant derivatives are defined as  \e\label{n4cov} D_\mu Z^A_d &=& \partial_\mu Z^A_d
-\tilde A_\mu{}^c{}_dZ^A_c ,\quad \tilde A_\mu{}^c{}_d=A^{ab}_\mu
f_{ab}{}^c{}_d+A^{\pa\pb}_\mu f_{\pa\pb}{}^c{}_d ,\\ \nonumber D_\mu Z^\DA_\pd &=&
\partial_\mu Z^\DA_\pd -\tilde A_\mu{}^\pc{}_\pd Z^\DA_\pc ,\quad \tilde
A_\mu{}^\pc{}_\pd=A^{\pa\pb}_\mu f_{\pa\pb}{}^\pc{}_\pd+A^{ab}_\mu f_{ab}{}^\pc{}_\pd. \ee
Here $A^{ab}_\mu$ and $A^{\pa\pb}_\mu$ are independent Hermitian tensors. In this way, we
have gauged the symmetry associated with the transformation (\ref{glbtran1}), i.e. the
symmetry generated  by the four subsets of 3-brackets (\ref{own3brcks}) and (\ref{mbrcks})
and  the four subsets of FIs (\ref{FI4}). The FIs (\ref{FI4}) can be also derived by
requiring that all structure constants are gauge invariant quantities, i.e. \e D_\mu
f_{abcd}=D_\mu f_{ab\pc\pd}=D_\mu f_{\pa\pb\pc\pd}=0. \ee

The $\CN=4$ Lagrangian, derived from a superspace approach, is given by
\e\label{LN4}\CL&=&\frac{1}{2}(-D_\mu\bar{Z}^A_aD^\mu Z^a_A-D_\mu\bar{Z}^\DA_\pa D^\mu
Z^\pa_\DA+i\bp^\DA_a\g^\mu D_\mu\p^a_\DA+i\bp^A_\pa\g^\mu
D_\mu\p^\pa_A)\nonumber\\&&-\frac{i}{2}(f_{acbd}Z^a_AZ^{Ab}\p^c_\DB\p^{\DB d}+
f_{\pa\pc\pb\pd}Z^\pa_\DA Z^{\DA\pb}\p^\pc_B\p^{B\pd })\nonumber\\&&+\frac{i}{2}
f_{ab\pc\pd}(Z^a_AZ^b_B\p^{A\pc}\p^{ B\pd}+Z^{\pc}_\DA Z^{\pd}_\DB\p^{\DA a}\p^{\DB b}+4Z^a_A
Z^{\DB\pd}\p^b_\DB\p^{A\pc})\nonumber\\&&
+\frac{1}{2}\epsilon^{\mu\nu\lambda}(f_{abcd}A_\mu^{ab}\partial_\nu
A_\lambda^{cd}+\frac{2}{3}f_{abc}{}^gf_{gdef}A_\mu^{ab}A_\nu^{cd}A_\lambda^{ef})\nonumber\\
&&+\frac{1}{2}\epsilon^{\mu\nu\lambda}(f_{\pa\pb\pc\pd}A_\mu^{\pa\pb}\partial_\nu
A_\lambda^{\pc\pd}+\frac{2}{3}f_{\pa\pb\pc}{}^\pg
f_{\pg\pd\pe\prf}A_\mu^{\pa\pb}A_\nu^{\pc\pd}A_\lambda^{\pe\prf})\nonumber\\
&&+\epsilon^{\mu\nu\lambda}(f_{ab\pc\pd}A_\mu^{ab}\partial_\nu A_\lambda^{\pc\pd}+f_{abc}{}^g
f_{gd\pe\prf}A_\mu^{ab}A_\nu^{cd}A_\lambda^{\pe\prf}+f_{ab\pc}{}^\pg
f_{\pg\pd\pe\prf}A_\mu^{ab}A_\nu^{\pc\pd}A_\lambda^{\pe\prf})\nonumber\\
&&+\frac{1}{12}(f_{abcg}f^g{}_{def}Z^{Aa}Z^b_BZ^{B(c}Z^{d)}_CZ^{Ce}Z^f_A
+f_{\pa\pb\pc\pg}f^\pg{}_{\pd\pe\prf}Z^{\DA\pa}Z^\pb_\DB Z^{\DB(\pc}Z^{\pd)}_\DC
Z^{\DC\pe}Z^\prf_\DA)\nonumber\\ &&-\frac{1}{4}(f_{ab \pc\pg}f^\pg{}_{\pd
ef}Z^{\DA\pc}Z^\pd_\DA Z^b_D Z^{Df}Z^a_C Z^{Ce}+f_{\pa\pb
cg}f^g{}_{d\pe\prf}Z^{Ac}Z^d_AZ^\pb_\DD Z^{\DD\prf}Z^\pa_\DC Z^{\DC\pe}).\nonumber\\ \ee %
And the $\CN=4$ supersymmetry transformations are  \e \label{SUSY4}&&\delta
Z^a_A=i\ep_A{}^\DA\p^a_\DA,\nonumber\\ &&\delta
Z^\pa_\DA=i\ep^\dag_\DA{}^A\p^\pa_A,\nonumber\\ &&\delta\p^\pa_A=-\g^\mu D_\mu
Z^\pa_\DB\ep_A{}^\DB-\frac{1}{3}f^\pa{}_{\pb\pc\pd}Z^\pb_\DB
Z^{\DB\pc}Z^\pd_\DC\ep_A{}^\DC+f^\pa{}_{\pb cd}Z^\pb_\DA Z^{Bc}Z^d_A\ep_B{}^\DA, \nonumber\\
&&\delta\p^a_\DA=-\g^\mu D_\mu Z^a_B\ep^\dag_\DA{}^B-\frac{1}{3}f^a{}_{bcd}Z^b_B
Z^{Bc}Z^d_C\ep^\dag_\DA{}^C+f^a{}_{b \pc\pd}Z^b_A
Z^{\DB\pc}Z^\pd_\DA\ep^\dag_\DB{}^A,\nonumber\\ &&\delta\tilde
A_\mu{}^c{}_d=i\ep^{A\DB}\g_\mu\p^b_\DB Z^a_Af_{ab}{}^c{}_d+i\ep^{\dag\DA
B}\g_\mu\p^\pb_BZ^\pa_\DA f_{\pa\pb}{}^c{}_d,\nonumber\\ &&\delta\tilde
A_\mu{}^\pc{}_\pd=i\ep^{A\DB}\g_\mu\p^b_\DB Z^a_Af_{ab}{}^\pc{}_\pd+i\ep^{\dag\DA
B}\g_\mu\p^\pb_BZ^\pa_\DA f_{\pa\pb}{}^\pc{}_\pd.\ee 
$\ep_{A}{}^{\dot{B}}$ satisfies the reality condition \begin{equation}\label{n4para}
\ep^{\dag}{}_{\dot{A}}{}^{B}= -\epsilon^{BC}\epsilon_{\dot{A}\dot{B}}\ep_{C}{}^{\dot{B}}.
\end{equation} We have explicitly verified the closure of the above $\CN=4$ superalgebra
\cite{ChenWu3}.

If the twisted and untwisted multiplets take values in the \emph{same}
symplectic 3-algebra, for instance, $\Phi_A=\Phi^a_A T_a$ and $\Phi_{\dot A}=\Phi^a_{\dot A}T_a$, the $\CN=4$ supersymmetry can be promoted to $\CN=5$ (see section \ref{secGN4}).

\section{Superalgebra Realization}\label{srt0} 
In this section, we will first demonstrate how to use \emph{two} superalgebras to construct
the \emph{four} sets of 3-brackets (\ref{own3brcks}) and (\ref{mbrcks}) and the \emph{four}
sets of FIs (\ref{FI4}); we will then comment on the two other 3-brackets (\ref{mbrcks3}) and
the related four sets of FIs (\ref{FI5}), though we do not really need (\ref{mbrcks3}) and
(\ref{FI5}) in constructing the theories.

Let us first briefly review the superalgebra construction of the symplectic 3-algebra in the
$\CN=5$ theory \cite{ChenWu3}. In the $\CN=5$ case, we have used the following superalgebra
\e\label{SLie} [M^m, M^n]=C^{mn}{}_sM^s,\quad [M^m,Q_R]=-\t^m_{RS}\omega^{ST}Q_T,\quad \{Q_R,Q_S\}=\t^m_{RS}k_{mn}M^n,\ee
to realize the symplectic 3-algebra. Here $R=1,\cdots, 2L$, and $\omega^{ST}=-\omega^{TS}$ and $k_{mn}$ are
invariant quadratic forms on the superalgebra. The key idea of the superalgebra realization
of 3-algebra is to identify the 3-algebra generators $T_R$ with the fermionic generators
$Q_R$, and to construct the 3-brackets in terms of double graded commutators on the
superalgebra, i.e., \begin{equation}\label{dcom} T_R\doteq Q_R, \quad [T_R, T_S; T_T]\doteq
[\{Q_R, Q_S\}, Q_T]. \end{equation} In this realization, the FI of the 3-algebra can be
converted into the $MMQ$ Jacobi identity of the superalgebra, and the constraint equation
$f_{(RST)U}=0$ for enhancing the $\CN=1$ supersymmetry to $\CN=5$ is equivalent to the $QQQ$
Jacobi identity of the superalgebra. The resulting Lie algebra of the gauge group is just the
bosonic subalgebra of the superalgebra (\ref{SLie}), and the corresponding representation is
determined by the fermionic generators.

As for the $\CN=4$ case, since both the double-symplectic 3-algebra and the $\CN=4$ three
algebra contain \emph{two} sub symplectic 3-algebras taking the same form as the sympletic
3-algebra in $\CN=5$ theory, it is natural to introduce \emph{two} superalgebras $G$ and
$G^\prime$, taking the same form as (\ref{SLie}), to construct the \emph{two} sub symplectic
3-algebras, respectively.  Hence the superalgebras $G$ and $G^\prime$ are given by
\begin{equation}\label{slie2} [M^u, M^v]=f^{uv}{}_wM^w,\quad [M^u,
Q_a]=-\t^u_{ab}\omega^{bc}Q_c,\quad \{Q_a,Q_b\}=\t^u_{ab}k_{uv}M^v, \end{equation} and
\begin{equation}\label{slie3} [M^\pu, M^\pv]=f^{\pu\pv}{}_\pw M^\pw,\quad [M^\pu,
Q_\pa]=-\t^\pu_{\pa\pb}\omega^{\pb\pc}Q_\pc,\quad
\{Q_\pa,Q_\pb\}=\t^\pu_{\pa\pb}k_{\pu\pv}M^\pv, \end{equation} respectively. Here
$a=1,\cdots,2R$ and $\pa=1,\cdots, 2S$; $\omega^{ab}$ and $\omega^{\pa\pb}$ are invariant
antisymmetric tensors whose inverses are denoted as $\omega_{bc}$ and $\omega_{\pb\pc}$,
satisfying $\omega^{ab}\omega_{bc}=\d^a_c$ and $\omega^{\pa\pb}\omega_{\pb\pc}=\d^\pa_\pc$;
$k_{uv}$ and $k_{\pu\pv}$ are invariant symmetric forms whose inverses are denoted as
$k^{vw}$ and $k_{\pv\pw}$, satisfying $k_{uv}k^{vw}=\d^w_u$ and
$k_{\pu\pv}k^{\pv\pw}=\d^\pw_\pu$. More explicitly, these bilinear forms are defined as
\cite{FSS} \begin{equation}\label{forms} \omega_{ab}=\kappa(Q_a,Q_b),\quad
\omega_{\pa\pb}=\kappa(Q_\pa,Q_\pb),\quad k^{uv}=-\kappa(M^u,M^v),\quad
k^{\pu\pv}=-\kappa(M^\pu,M^\pv), \end{equation} which are invariant \cite{GaWi,FSS} in the
sense that \begin{equation}\label{compatible0} \kappa([A,B\},C)=\kappa(A,[B,C\}),\quad
\kappa([A^\prime,B^\prime\},C^\prime)=\kappa(A^\prime,[B^\prime,C^\prime\}), \end{equation}
where $A=Q_a$ \emph{or} $M^u$, and $A^\prime=Q_\pa$ \emph{or} $M^\pu$. For example,
\begin{equation}\label{compatible} \kappa(\{Q_a,Q_b\},M^v)=\kappa(Q_a,[Q_b,M^v]),\quad
\kappa([M^u,M^v],M^w)=\kappa(M^u,[M^v,M^w]). \end{equation} (The minus sign on the RHS of the
third equation of (\ref{forms}) is determined by the first equation of (\ref{compatible}) and
the convention that $\t^v_{ab}=\omega_{ac}\t^{vc}{}_b$.)

In analogue to (\ref{dcom}), we construct the 3-brackets in (\ref{own3brcks}) as follows
\e\label{4dgdcm} &&T_a\doteq Q_a, \quad [T_a, T_b; T_c]\doteq [\{Q_a, Q_b\},
Q_c]=f_{abc}{}^dQ_d, \\ &&T_\pa\doteq Q_\pa, \quad [T_\pa, T_\pb; T_\pc]\doteq [\{Q_\pa,
Q_\pb\}, Q_\pc]=f_{\pa\pb\pc}{}^\pd Q_\pd.\nonumber \ee We note immediately that
\e\label{subsa} f_{abc}{}^\pd=f_{\pa\pb\pc}{}^d=0, \ee i.e. Eqs (\ref{vanish}) are indeed
satisfied.

As we introduced in Section \ref{Introduction}, it is natural to construct the two 3-brackets
(\ref{mbrcks}) in terms of the two double graded commutators on $G$ \emph{and} $G^\prime$:
\begin{equation}\label{mbrcks2} [T_a,T_b;T_\pc]\doteq [\{Q_a, Q_b\}, Q_\pc], \quad
[T_\pa,T_\pb;T_c]\doteq [\{Q_\pa, Q_\pb\}, Q_c]. \end{equation} They must satisfy two crucial
conditions. First, there is a \emph{physical requirement}: To guarantee non-trivial
interactions between the twisted and untwisted multiplets, we must require that
\e\label{2p2u} f_{ab\pc}{}^\pd\neq0,\quad f_{\pa\pb c}{}^d\neq0. \ee Secondly, we require that
\e\label{vanish2} f_{ab\pc}{}^d=f_{\pa\pb c}{}^\pd =0. \ee (See Eqs. (\ref{unused0}).) We
will demonstrate that the above two conditions can be satisfied simultaneously by imposing
certain conditions on the bosonic parts of $G$ and $G^\prime$; then we will be able to prove
Eqs. (\ref{ortho}): \begin{equation}\label{ortho2} \omega(T_a, T_\pb)\doteq\kappa(Q_a,
Q_\pb)=0,\quad\omega(T_\pb, T_a)\doteq\kappa(Q_\pb, Q_a)=0. \end{equation}

Let us first
examine $[\{Q_a,Q_b\},Q_\pc]$. Using (\ref{slie2}), a short computation gives
\begin{equation}\label{mxdgcm} [T_a,T_b;T_\pc]\doteq[\{Q_a,Q_b\},Q_\pc]=k_{uv}\t^u_{ab}[M^v,
Q_\pc]. \end{equation} So requiring that $f_{ab\pc}{}^\pd\neq 0$ and $f_{ab\pc}{}^d=0$ is
equivalent to requiring that \e\label{cmmfct} [M^v, Q_\pc]=\t^{v\pd}{}_\pc Q_\pd \ee with
$\t^{v\pd}{}_\pc\neq 0$, which means that the set of fermionic generators $Q_\pc$ furnish a
nontrivial representation of $M^v$.

It is sufficient for that $\t^{v\pd}{}_\pc\neq 0$ if the Lie algebra
spanned by $M^v$ share at least one simple factor or $U(1)$ factor with the Lie algebra
spanned by $M^{\pv}$. To prove this statement, we denote the generators of the common bosonic
part as $M^g$, i.e. schematically, \e\label{cup} M^g=M^v\cap M^\pv. \ee (It is also allowed
that $M^g\subset M^v$, $\tilde M^g\subset M^\pv$ and $\tilde M^g=T^g{}_h M^h$ with $T^g{}_h$
an invertible complex matrix (see Section \ref{secosp4n} for an example).) Decompose $M^v$
and $M^\pv$ as \e\label{dcmm} M^v=(M^\a, M^g),\quad M^\pv=(M^{\a^\prime}, M^g), \ee where
$[M^\a, M^g]=[M^{\a^\prime}, M^g]=[M^\a, M^{\a^\prime}]=0$. (Here $\a$ is \emph{not} an index
of spacetime spinor.) 
And we assume that at least one of the two commutators $[M^\a, Q_a]$ and $[M^{\a^\prime}, Q_{a^\prime}]$ does not vanish, i.e. we exclude the possibility that
\e\label{nonempty}
[M^\a, Q_a]=[M^{\a^\prime}, Q_{a^\prime}]=0.
\ee
We can further sharpen Eqs. (\ref{cup}) and
(\ref{dcmm}) by requiring that \e\label{cmmt} [M^{\a^\prime},Q_a]=0,\quad [M^\a,Q_\pa]=0. \ee

Now we assume that the superalgebra $G^\prime$ is simple\footnote{The definition of a simple
superalgebra is analogous to that of a simple Lie algebra: A simple superalgebra is a
superalgebra without any invariant proper sub-superalgebras. A sub-superalgebra $\cal{I}$ is
called invariant if the commutator or anti-commutator of any generator of the whole
superalgebra $\cal{S}$ with any generator of the sub-superalgebra is still in $\cal{I}$,
i.e., $[X,Y\}\subset \cal{I}$, $(X \subset {\cal S}, Y\subset\cal{I})$.}. If $[M^g,
Q_\pc]=0$, then the Lie algebra defined by $M^g$ must be an invariant subalgebra of the
superalgebra $G^\prime$, which contradicts our assumption that $G^\prime$ is a simple
superalgebra. We therefore must always have \e\label{cmfct} [M^v, Q_\pc]=[M^g,
Q_\pc]=\t^{g\pd}{}_\pc Q_\pd, \quad \t^{g\pd}{}_\pc\neq 0. \ee If $G^\prime$ is \emph{not}
simple, the right hand side of Eq. (\ref{cmmfct}) still does not vanish, provided that the
common part $M^g$ is \emph{not} the center of $G^\prime$.

Similarly, we can demonstrate that
the second equation of (\ref{2p2u}) and the second equation of (\ref{vanish2}) are obeyed
under the same conditions imposed on $G$ and $G^\prime$. Now Eqs. (\ref{mbrcks2}) become
\e\label{mbrcks20} &&[T_a,T_b;T_\pc]\doteq [\{Q_a, Q_b\}, Q_\pc]=f_{ab\pc}{}^\pd
Q_\pd,\nonumber \\ &&[T_\pa,T_\pb;T_c]\doteq [\{Q_\pa, Q_\pb\}, Q_c]=f_{\pa\pb c}{}^dQ_d. \ee
We see that the conditions (\ref{2p2u}) and (\ref{vanish2}) are satisfied.

We now want to
prove that Eqs. (\ref{ortho2}) are also satisfied. Consider the following equation \e
\kappa([\{Q_a,Q_b\},Q_c],Q_\pd)=\kappa([\{Q_a,Q_b\},Q_\pd],Q_c). \ee A short computation
gives \e\label{schur} k_{uv}\t^u_{ab}(\t^v)^d{}_c\omega_{d\pd}=(k_{\a\b}\t^\a_{ab}\t^{\b
d}{}_c+k_{gh}\t^g_{ab}(\t^h)^d{}_c)\omega_{d\pd} =k_{gh}\t^g_{ab}(\t^h)^\pc{}_\pd \omega_{\pc
c}, \ee where $\omega_{d\pd}=\kappa(Q_d,Q_\pd)$. We have decomposed $k_{uv}$ into
$k_{uv}=(k_{\a\b}, k_{gh})$; in the most right hand-side, we have used the second equation of (\ref{cmmt}). Raising the index $c$, the above equation can be written as
\e\label{schur2} k_{uv}\t^u_{ab}(\t^v)^c{}_d\omega^d{}_\pd=(k_{\a\b}\t^\a_{ab}\t^{\b
c}{}_d+k_{gh}\t^g_{ab}(\t^h)^c{}_d)\omega^d{}_\pd=\omega^c{}_\pc(k_{gh}\t^g_{ab}(\t^h)^\pc{}_\pd),
\ee where $\omega^d{}_\pd=\omega^{de}\omega_{e\pd}$.

Recall that we exclude the possibility that the two commutators $[M^\a, Q_a]$ and $[M^{\a^\prime}, Q_{a^\prime}]$ vanish identically (see Eqs (\ref{nonempty})).  Without loss generality, we assume that
$[M^\a, Q_c]=(\t^\a)^d{}_cQ_d\neq0$. In other words,
$(\t^\a)^d{}_c\neq0$, i.e. it is a nontrivial (and irreducible)
representation of $M^\a$ furnished by the fermionic generators $Q_d$.
However, the right hand side of (\ref{schur}) indicates that $Q_\pd$
furnish a trivial representation of $M^\a$ in the sense that $(\t^\a)^\pc{}_\pd=0$.
In summary, $k_{uv}\t^u_{ab}(\t^v)^d{}_c$ and $k_{gh}\t^g_{ab}(\t^h)^\pc{}_\pd$ are nonequivalent
 and irreducible representations\footnote{Generally speaking, the set of
 generators $Q_c=Q_{\bk\hk}$  furnish a bi-fundamental representation of the anticommutator
 $\{Q_a, Q_b\}=(\t_\a)_{ab}M^\a+(\t_g)_{ab}M^g$, in the sense that  $[\{Q_a, Q_b\}, Q_c]=(\t_\a)_{ab}(M^\a)^{\hat{l}}{}_{\hk} Q_{\bk\hat{l}}+(\t_g)_{ab}(M^g)^{\bar{l}}{}_{\bk} Q_{\bar{l}\hk}$. Here $\bk$ and $\hk$ are fundamental indices of $M^g$ and $M^\a$, respectively. The two Lie algebras,
 spanned by $M^\a$ and $M^g$ respectively, are \emph{distinct}
 and have nothing in common;
 the fundamental indices $\bk$ and $\hk$
 are \emph{distinct} as well.
 So this bi-fundamental representation is irreducible.
 On the other hand, since $[M^\a,Q_\pd]=0$, $Q_\pd$ only furnish an irreducible representation of $(\t_g)_{ab}M^g$. (See section \ref{secNewEx} and section \ref{Allex} for many examples.)} of $\{Q_a, Q_b\}$ furnished by $Q_c$ and $Q_\pd$, respectively. Apply Schur's Lemma to  equation (\ref{schur2}), we have immediately $\omega^d{}_\pd=0$. Hence $\omega_{d\pd}=0$ on account of that $\omega^{de}$ is nonsingular. So Eqs. (\ref{ortho2}) are satisfied.


Substituting (\ref{cmmfct}) into the first equation of
(\ref{mbrcks2}) gives the structure constants
\e\label{tensorprd2} f_{ab\pc\pd}=
k_{gh}\t^g_{ab}\t^h_{\pc\pd}, \ee
Using (\ref{slie2}) and
(\ref{slie3}), a short calculation gives the structure constants
of 3-brackets in (\ref{4dgdcm})
\e\label{tensorprd1}
f_{abcd}=k_{uv}\t^u_{ab}\t^v_{cd}, \quad
f_{\pa\pb\pc\pd}=k_{\pu\pv}\t^\pu_{\pa\pb}\t^\pv_{\pc\pd}. \ee
The structure constants (\ref{tensorprd2}) and (\ref{tensorprd1}) possess the desired reality and symmetry
properties. The $Q_aQ_bQ_c$ ($Q_\pa Q_\pb Q_\pc$) Jacobi identity
implies that $f_{(abc)d}=0$ ($f_{(\pa\pb\pc)\pd}=0$) guaranteeing
that the supersymmetry can be enhanced from $\CN=1$ to $\CN=4$.

The four sets of FIs in (\ref{FI4}) are equivalent to the
$M^uM^vQ_a$, $M^uM^vQ_\pa$, $M^\pu M^\pv Q_a$ and $M^\pu M^\pv
Q_\pa$ Jacobi identities, respectively. For
instance, by using Eqs. (\ref{4dgdcm}) and (\ref{mbrcks20}), it is not
difficult to prove that the following equation (as one of equations (\ref{FI}))
\begin{eqnarray}\label{n4FISP}
[\{Q_a,Q_b\},[\{Q_c,Q_d\},Q_\pa]]&=&[\{[\{Q_a,Q_b\},Q_c],Q_d\},
Q_\pa]+[\{Q_c,[\{Q_a,Q_b\},Q_d]\},
Q_\pa]\nonumber\\&&+[\{Q_c,Q_d\},[\{Q_a,Q_b\},Q_\pa]]
\end{eqnarray}
is equivalent to the second FI of (\ref{FI4}). On the other hand,
using the superalgebras (\ref{slie2}) and (\ref{slie3}), one can
easily convert the above equation into the $M^uM^vQ_\pa$ Jacobi
identity
\e\label{n4MMQjcb}
\t^u_{ab}\t^v_{cd}([M_v,[M_u,Q_\pa]]-[M_u,[M_v,Q_\pa]]+[[M_u,
M_v],Q_\pa])=0,\ee
where we have used the equation \e\label{totanti}
k_{uw}k_{vx}f^{xw}{}_y+k_{uw}k_{yx}f^{xw}{}_v=0, \ee
which is equivalent to the second equation of (\ref{compatible}) on account of that
$k^{uv}$ is invertible.
Therefore the
second FI of (\ref{FI4}) is equivalent to the $M^uM^vQ_\pa$ Jacobi
identity.
Using Eqs. (\ref{cup}) and (\ref{dcmm}), Eq. (\ref{n4MMQjcb}) becomes
\e\label{n4MMQjcb2}
\t^g_{ab}\t^h_{cd}([M_h,[M_g,Q_\pa]]-[M_g,[M_h,Q_\pa]]+[[M_g,
M_h],Q_\pa])=0,\ee
which is of course obeyed, since the $M^\pu M^\pv Q_\pa$ Jacobi
identity is obeyed and $M^g\subseteq M^\pu$.

In this realization, the un-twisted and twisted multiplets take values in the bosonic
subalgebras of the superalgebras (\ref{slie2}) and (\ref{slie3}),
respectively; and the representations of the bosonic parts
(\ref{slie2}) and (\ref{slie3}) are determined by the fermionic
 generators $Q_a$ and $Q_\pa$, respectively.

Here we have to emphasize that so far we have constructed only the \emph{four} structure constants of the 3-brackets (\ref{own3brcks}) and (\ref{mbrcks}) in terms of tensor products (\ref{tensorprd1}) and (\ref{tensorprd2}) on the superalgebras $G$ and $G^\prime$, respectively, and solved the four sets of
FIs (\ref{FI4}) of the double-symplectic 3-algebra in terms of certain Jacobi identities of the superalgebras $G$ and $G^\prime$. These constructions are sufficient for the purpose of classifying the gauge groups of the $\CN=4$ theory (see Section (\ref{Allex})).


We now would like to comment on the rest two 3-brackets
(\ref{mbrcks3}) and the rest four FIs (\ref{FI5}), though they do not play any role in constructing the theories. In the case of double-symplectic 3-algebra, we will demonstrate that at least in the special case of that $G$ and $G^\prime$ can be `fused' into a single closed superalgebra (the fusion procedure will be introduced in a
separated paper \cite{ChenWu5}), Eqs. (\ref{mbrcks3}) and (\ref{FI5}) can be constructed in terms of superalgebras as well.

In the case of double-symplectic 3-algebra, under the condition that $G$ and $G^\prime$ can be `fused' into a closed superalgebra, the rest two 3-brackets Eqs. (\ref{mbrcks3}) can be constructed in analogue to Eqs. (\ref{4dgdcm}) and (\ref{mbrcks2}), i.e.
\e\label{unused3b}
[T_a,T_\pb;T_c]\doteq[\{Q_a, Q_\pb\}, Q_c],\quad
[T_\pa,T_b;T_\pc]\doteq[\{Q_\pa, Q_b\}, Q_\pc],
\ee
where we have use the fact that the disjoint union of the two sets generators $Q_a$ and $Q_\pa$ form a complete fermionic basis of the ``fused" superalgebra \cite{ChenWu5}. In summary, we have
\e\label{alldb}
T_I\doteq Q_I,\quad[T_I,T_J;T_K]\doteq [\{Q_I, Q_J\}, Q_K],
\ee
where the index $I=a$ \emph{or} $\pa$. Using the first equation of Eq. (\ref{compatible}), it is easy to prove that
\e\label{consistent}
\kappa([\{Q_a,Q_b\},Q_c],Q_d)=\kappa(\{Q_a,Q_b\},\{Q_c,Q_d\}).
\ee
Generally speaking, the following equations
\e\label{consistentg}
\kappa([\{Q_I,Q_J\},Q_K],Q_L)=\kappa(\{Q_K,Q_L\},\{Q_I,Q_J\})
=\kappa([\{Q_K,Q_L\},Q_I],Q_J)
\ee
hold. The above equations imply that (\ref{bigsym}) is obeyed by the construction
\e\label{bigsym2}
\omega([T_I,T_J;T_K],T_L)\doteq\kappa([\{Q_I,Q_J\},Q_K],Q_L).
\ee
Using (\ref{subsa}), (\ref{consistentg}) and (\ref{ortho2}), we obtain $f_{a\pb c}{}^d=f_{\pa b \pc}{}^\pd=0$. In summary, we have
\e\label{subsa3}
f_{a\pb c}{}^d=f_{\pa b \pc}{}^\pd=f_{ab\pc}{}^d=f_{\pa\pb c}{}^\pd=f_{abc}{}^\pd=f_{\pa\pb\pc}{}^d=0,
\ee
which are nothing but Eqs. (\ref{unused0}) and (\ref{vanish}). Combing (\ref{consistentg}), (\ref{bigsym2}) and (\ref{subsa3}), one can prove that (\ref{bigsym1}) is also satisfied. The first two equations of Eqs. (\ref{subsa3}), i.e. $f_{a\pb c}{}^d=f_{\pa b \pc}{}^\pd=0$, imply that the two structure constants associated with the brackets (\ref{unused3b}) are given by
\e\label{mbrcks4} &&[T_a,T_\pb;T_c]\doteq[\{Q_a,Q_\pb\},
Q_c]=f_{a\pb c}{}^\pd Q_\pd,\nonumber\\&& [T_\pa,T_b;T_\pc]\doteq[\{Q_\pa,Q_b\}, Q_\pc]=f_{\pa b\pc}{}^dQ_d.
\ee
Define the anticommutator of $Q_a$ and $Q_\pb$ as
\e\label{mtm}
\{Q_a,Q_\pb\}=t^{\tu}_{a\pb}M_{\tu},
\ee
where $M_{\tu}$ are a set of bosonic generators. We have to emphasis that we have not introduced any new generators into the ``fused" superalgebra except  $M^{\tilde u}$ \cite{ChenWu5}. So the generators of the ``fused" superalgebra are consisted of by the generators of $G$ and $G^\prime$, as well as $M_{\tu}$. \emph{Every} Jacobi identity of the ``fused" superalgebra is obeyed.   In particular, the Jacobi identities
\e\label{JI1}
&&[\{Q_a,Q_\pb\}, Q_c]+[\{Q_a,Q_c\}, Q_\pb]+[\{Q_c,Q_\pb\}, Q_a]=0,\\
&&[\{Q_\pa,Q_b\}, Q_\pc]+[\{Q_\pa,Q_\pc\}, Q_b]+[\{Q_\pc,Q_b\}, Q_\pa]=0\label{JI2}
\ee
are satisfied. In other words, we have the identity
\begin{equation}\label{JI}
f_{a\pb c\pd}+f_{c\pb a\pd}+f_{ac\pb \pd}=0\quad{\rm or}\quad\t^{\tilde u}_{a\pb}(\t_{\tilde u})_{c\pd}+\t^{\tilde u}_{c\pb}(\t_{\tilde u})_{a\pd}
+\t^g_{ac}(\t_g)_{\pb\pd}=0,
\end{equation}
which follows from (\ref{mtm}) via (\ref{JI1}) or (\ref{JI2}). Therefore these structure constants (or the corresponding 3-brackets) are \emph{not} independent in this special case.

Also, under the condition that
$G$ and $G^\prime$ can be `fused' into a single closed
superalgebra, it is not difficult to prove that the four FIs (\ref{FI5}) involving $f_{a\pb c\pd}$ can be converted into certain Jacobi identities of the ``fused" superalgebra. These Jacobi identities involve the bosonic generators $M_{\tu}$ defined in Eq. (\ref{mtm}). In this way, the whole double-symplectic 3-algebra can be realized in terms of ``fused" superalgebras; however, we are not sure whether it can be realized in terms of $G$ and $G^\prime$ or not, if $G$ and $G^\prime$ cannot be `fused' into a single closed superalgebra. We therefore leave it as an open question.


In the case of $\CN=4$ three-algebra, the two structure constants $f_{a\pb c}{}^\pd$ and $f_{a\pb\pc}{}^d$, vanishing identically, \emph{cannot} be constructed in terms of the tensor products on the superalgebras $G$ and $G^\prime$, since the double graded commutators $[\{Q_a, Q_\pb\}, Q_c]=f_{a\pb c}{}^\pd Q_\pd$ and $[\{Q_a, Q_\pb\}, Q_\pc]=f_{a\pb \pc}{}^d Q_d$ do \emph{not} vanish. For example, if both $G$ and $G^\prime$ are orthosymplectic superalgebras \emph{or} unitary superalgebras, one can prove that both $[\{Q_a, Q_\pb\}, Q_c]$ and $[\{Q_a, Q_\pb\}, Q_\pc]$ are \emph{not} zero by direct calculation \cite{ChenWu5}.



\section{Explicit Examples of New $\CN=4$ Quiver Theories}\label{secNewEx}
The classification of gauge groups of the $\CN=4$ theories can be
found in \cite{HosomichiJD}, \cite{Bergshoeff} (there is a nice
summary in \cite{MFM:Aug09}). In this section, however, we are
able to construct some new examples which were neglected in
previous classification in the literature by using the ideas
described in the previous section.

\subsection{$Sp(2N)\times U(1)\times U(M)$}\label{secospunu1}
Here we choose the superalgebras $G$ and $G^\prime$ (see
(\ref{slie2}) and (\ref{slie3})) as $OSp(2|2N)$ and $U(M|1)$
respectively. The common part of the bosonic parts of $OSp(2|2N)$
and $U(M|1)$ is $SO(2)\cong U(1)$. Some useful $U(M|1)$
commutation relations are (the commutation relations of $U(M|N)$
are given by Appendix \ref{cmrunum}) \e\label{cmrunum2}
&&\{\bar{Q}_\pri, Q^\pj\}=k^\prime(M_\pri{}^\pj+\d_\pri{}^\pj
M_{U(1)}),
\quad [M_{U(1)}, \bar{Q}_\pri]=-\bar{Q}_\pri,\quad [M_{U(1)}, Q^\pri]=Q^\pri,\nonumber\\
&& [M_\pri{}^\pj, \bar{Q}_\pk]=\d_\pk{}^\pj \bar{Q}_\pri,\quad
[M_\pri{}^\pj, Q^\pk]=-\d_\pri{}^\pk Q^\pj, \ee where the
subscript index $\pri=1,\ldots M$ is the index for the fundamental
representation of $U(M)$, and $M_{U(1)}$ is the $U(1)$ generator.
We have suppressed the $U(1)$ indices carried by the fermionic
generators.
We then identify the $U(1)$ generator $M_{U(1)}$ with the $SO(2)$
generator of $OSp(2|2N)$ $M_{\bi\bj}$ (the commutation relations
of $OSp(M|2N)$ are given by Appendix \ref{cmrsp22n}) by imposing
the following equation \e\label{u1eso2} \epsilon_{\bi\bj}M_{U(1)}=
-M_{\bi\bj}. \ee Here $\epsilon_{\bi\bj}=-\epsilon_{\bj\bi}$ and
$\epsilon_{\bar 1\bar 2}=1$, with $\bar i=1,2$ an $SO(2)$ index.
With this identification, the commutator of $M_{U(1)}$ with a
fermionic generator of $OSp(2|2N)$ is given by \e [M_{U(1)},
Q_{\bi\hi}]=\ep_{\bi\bj}Q_{\bj\hi}; \ee and the commutators of the
$SO(2)$ generator of $OSp(2|2N)$ $M_{\bi\bj}$ with the fermionic
generators of $U(M|1)$ are given by \e\label{onion} [M_{\bi\bj},
\bar{Q}_\pri]=\ep_{\bi\bj}\bar{Q}_\pri, \quad [M_{\bi\bj},
Q^\pri]=-\ep_{\bi\bj}Q^\pri. \ee To calculate the structure
constants, we define \e Q_a=Q_{\bi\hi}\quad {\rm and} \quad
Q_\pa=\begin{pmatrix}
\bar{Q}_\pri\\-Q^\pri\end{pmatrix}=\bar{Q}_\pri\d_{1\a}-Q^\pri\d_{2\a},
\ee where $\hi=1,\cdots, 2M$ is an $Sp(2N)$ fundamental index.  In
the second equation we have introduced a ``spin up" spinor
$\chi_{1\alpha}=\d_{1\a}$ and a ``spin down" spinor
$\chi_{2\alpha}=\d_{2\a}$ \footnote{Here $\a$ is \emph{not} a
spacetime spinor index. In this paper, since we have to label many
indices, it is unavoidable that some letters will be repeatedly
used, but they will be defined explicitly in the sections in which
they are used. We hope this will not cause any confusion.}). The
structure constants $f_{ab\pc\pd}$ can be read from the double
graded commutator. \e [\{Q_a, Q_b\}, Q_\pc]=f_{ab\pc}{}^\pd Q_\pd.
\ee The double grade commutator can be calculated
straightforwardly by using (\ref{cmrunum2}), (\ref{onion}) and
Appendix \ref{cmrsp22n}. The structure constants $f_{ab\pc\pd}$
are given by \e\label{strct1}
f_{ab\pc\pd}=f_{\bi\hi,\bj\hj,\pc\pd}=k\omega_{\hi\hj}\epsilon_{\bi\bj}
(\d_\pri{}^\pj\d_{1\a}\d_{2\b}+\d_\pj{}^\pri\d_{1\b}\d_{2\a}). \ee
Similarly, one can calculate $f_{\pc\pd ab}$ by using \e [\{Q_\pc,
Q_\pd\}, Q_a]=f_{\pc\pd a}{}^b Q_b. \ee By requiring
$f_{ab\pc\pd}=f_{\pc\pd ab}$, $k^\prime$ in the anticommutator of
(\ref{cmrunum2}) is determined to be $k^\prime=-k$, with $k$
defined in the anticommutator in (\ref{OSp}).

The structure constants $f_{abcd}$ can be read off from the double
graded commutator $[\{Q_a, Q_b\}, Q_c]=f_{abc}{}^d Q_d$; they are
given by
\begin{eqnarray}\label{strct2}
f_{abcd}=f_{\bi\hat{i},\bj\hat{j},\bk\hat{k},\bl\hat{l}}
=k[(\delta_{\bi\bk}\delta_{\bj\bl}-\delta_{\bi\bl}\delta_{\bj\bk})
\omega_{\hat{i}\hat{j}}\omega_{\hat{k}\hat{l}}
-\delta_{\bi\bj}\delta_{\bk\bl}(\omega_{\hat{i}\hat{k}}
\omega_{\hat{j}\hat{l}}+\omega_{\hat{i}\hat{l}}
\omega_{\hat{j}\hat{k}})].
\end{eqnarray}
Similarly, we have
\begin{eqnarray}\label{strct3}
f_{\pa\pb\pc\pd}&=&f^{\pj\pl}{}_{\pk\pri}\d_{1\a}\d_{2\b}\d_{1\g}\d_{2\d}
+f^{\pk\pj}{}_{\pri\pl}\d_{1\a}\d_{2\b}\d_{2\g}\d_{1\d}\nonumber\\
&&+f^{\pri\pl}{}_{\pk\pj}\d_{2\a}\d_{1\b}\d_{1\g}\d_{2\d}
+f^{\pk\pri}{}_{\pj\pl}\d_{2\a}\d_{1\b}\d_{2\g}\d_{1\d},\\
f^{\pj\pl}{}_{\pk\pri}&\equiv& k(\d_\pk{}^\pj\d_\pri{}^\pl-\d_\pri{}^\pj\d_\pk{}^\pl)\nonumber.
\end{eqnarray}
It is straightforward to verify that the structure constants
(\ref{strct1}), (\ref{strct2}) and (\ref{strct3}) satisfy the
symmetry conditions (\ref{symfs}) and the reality conditions
(\ref{rltcndtn}), and obey the fundamental identities (\ref{FI4}).
Substituting (\ref{strct1}), (\ref{strct2}) and (\ref{strct3})
into (\ref{n4cov}), (\ref{LN4}) and (\ref{SUSY4}) gives the
$\CN=4, Sp(2N)\times U(1)\times U(M)$ theory. In this realization,
the un-twisted multiplets are in the bifundamental representation
of $Sp(2N)\times U(1)$, while the twisted multiplets are in the
bifundamental representation of $U(M)\times U(1)$. In Section
\ref{secospu}, we will introduce a more general scheme to identify
the $U(1)$ factors of the even parts of $OSp(2|N_2)$ and
$U(N_3|N_4)$. The resulting gauge group will be $Sp(2N_2)\times
U(1)\times U(N_3)\times U(N_4)$.


\subsection{$Sp(2N)\times SU(2)_R\times SU(2)_L\times SO(M)$ and
Other Examples}\label{secosp4n}

The superalgebras $OSp(4|2N)$ and $D(2|1,\alpha)$ (with $\a$ a
continuous parameter) are rather special for constructing $\CN=4$
theories in that both of them contain an $SO(4)$ factor in their
bosonic parts. A simple observation is the well known
decomposition $SO(4)\cong SU(2)_R\times SU(2)_L$, where the
$SU(2)_R$ ($SU(2)_L$) generators satisfy an anti-self-duality
(self-duality) condition. Since several classes of superalgebras
contain a simple factor $SU(2)$ (or its isometries $Sp(2)$ and
$SO(3)$) in their bosonic parts, identifying them with $SU(2)_R$
or $SU(2)_L$ in the bosonic parts of the superalgebra $OSp(4|2N)$
or $D(2|1,\alpha)$ will generate new $\CN=4$ theories.

In this subsection, we present an explicit example to illustrate
this idea.  We choose $G=OSp(4|2N)$ and $G^\prime=OSp(M|2)$. (The
commutation relations of $OSp(M|2N)$ are given by Appendix
\ref{cmrsp22n}.) The bosonic part of the superalgebra $OSp(4|2N)$
contains an $SO(4)$ factor. Later $SO(4)$ will be decomposed into
$SU(2)_R\times SU(2)_L$. The bosonic part of $OSp(M|2)$ includes
an $Sp(2)$ factor. It is well known that $Sp(2)\cong SU(2)$, so
without loss of generality, we can identify $Sp(2)$ with
$SU(2)_R$\footnote{Strictly speaking, the set of $SU(2)_R$
generators are related to the set of $Sp(2)$ generators via a
\emph{complex} linear nonsingular transformation (see Eqs
(\ref{2su2})).}. In other words, we choose \e\label{sp2esu2}
Sp(2)=SU(2)_R \ee as the common part of the bosonic parts of
$OSp(4|2N)$ and $OSp(M|2)$. Some useful commutation relations of
$OSp(M|2)$ are \e\label{cmrsp22n3}
&&\{Q_{\bi\a}, Q_{\bj\b}\}=k^\prime(\ep_{\a\b}M_{\bi\bj}
+2\d_{\bi\bj}M_{\a\b}),\quad [2M_{\a\b}, Q_{\bi\g}]
=\ep_{\b\g}Q_{\bi\a}+\ep_{\a\g}Q_{\bi\b},\nonumber\\
&&[M_{\bi\bj},Q_{\bk\a}]=\d_{\bj\bk}Q_{\bi\a}-\d_{\bi\bk}Q_{\bj\a},
\ee where in the first line we have introduced a factor $2$ for
later consistence. Here $\bi=1,\ldots, M$ is an $SO(M)$ index, and
$\a=1, 2$ is an $Sp(2)$ index, \emph{not} a spacetime spinor index
(we hope this will not cause any confusion).  And the useful
commutation relations of $OSp(4|N)$ are given by \e
&&\{Q_{m\hi}, Q_{n\hj}\}=k(\omega_{\hi\hj}M_{mn}+\d_{mn}M_{\hi\hj}),
\quad [M_{mn}, Q_{p\hi}]=\d_{np}Q_{m\hi}-\d_{mp}Q_{n\hi},\nonumber\\
&&[M_{\hi\hj},Q_{m\hk}]=\omega_{\hj\hk}Q_{m\hi}+\omega_{\hi\hk}Q_{m\hj}.
\ee
Here $m=1,\ldots, 4$ is an $SO(4)$ index, and $\hi=1,\ldots, 2N$
an $Sp(2N)$ index.

To decompose the $SO(4)$ generators, we first introduce a set of
$SU(2)_R\times SU(2)_L$ $\s$-matrices as follows
\e\label{paulim}
\sigma_{m\a}{}^{\dot\a}=(\sigma_1,\sigma_2,\sigma_3,
i\mathbb{I}),\quad
\sigma^{\dag}_{m\dot\a}{}^{\a}=(\sigma_1,\sigma_2,\sigma_3,
-i\mathbb{I}),
\ee%
where $\s_{i}$ ($i=1,\ldots, 3$) are pauli matrices. The $SU(2)_R$
matrices $\s_{mn}$ and the $SU(2)_L$ matrices $\bar\s_{mn}$ are
defined as \e\label{su2su2}
&&\s_{mn\a}{}^\b=\frac{1}{4}(\s_m\s^{\dag}_n-\s_n\s^{\dag}_m)_{\a}{}^\b,\quad
\bar\s_{mn\dot\a}{}^{\dot\b}=\frac{1}{4}(\s^{\dag}_m\s_{n}
-\s^{\dag}_n\s_{m})_{\dot\a}{}^{\dot\b}; \ee and they satisfy the
`duality' conditions
\begin{equation}\label{dltcdn1}
\s_{mn}=-\frac{1}{2}\vp_{mnpq}\s_{pq},\quad
\bar\s_{mn}=\frac{1}{2}\vp_{mnpq}\bar\s_{pq}.
\end{equation}
Here $\vp_{mnpq}$ is the totally antisymmetric
tensor.\footnote{Our convention is that $\vp^{1234}=1$ and
$\vp_{mnpq}=\vp^{mnpq}$.}
We decompose the $SO(4)$ generators $M_{mn}$ by defining
\e\label{mpm} M^{\pm}_{mn}=\frac{1}{2}(M_{mn}\pm
\frac{1}{2}\vp_{mnpq}M_{pq}). \ee With the above definitions, we
have \e\label{dcso4} M_{mn}=M^+_{mn}+M^-_{mn},\quad
M^{\pm}_{mn}=\pm\frac{1}{2}\vp_{mnpq}M^{\pm}_{pq}, \quad
[M^+_{mn}, M^-_{pq}]=0. \ee Therefore $M^+_{mn}$ and $M^-_{mn}$,
satisfying self-duality and anti-self-duality conditions, must be
the $SU(2)_L$ and $SU(2)_R$ generators, respectively. We now
connect the set of $Sp(2)$ generators $M_{\a\b}$ of $OSp(M|2)$
with $M^{-}_{mn}$ via the equation \footnote{ We use the invariant
antisymmetric tensor $\ep_{\a\b}$ and $\ep_{\dot\a\dot\b}$ to
lower undotted and dotted indices, respectively. For example,
$\sigma^{\dag}_{m\dot\a\g}=\ep_{\b\g}\sigma^{\dag}_{m\dot\a}{}^{\g}$
and
$\sigma_{m\a\dot\g}=\ep_{\dot\b\dot\g}\sigma_{m\a}{}^{\dot\g}$.
The inverse of $\ep_{\a\b}$ is defined as $\ep^{\b\g}$ satisfying
$\ep_{\a\b}\ep^{\b\g}=\d_{\a}{}^{\g}$ (similarly
$\ep_{\dot\a\dot\b}\ep^{\dot\b\dot\g}=\d_{\dot\a}{}^{\dot\g}$).}
\e\label{2su2} M^-_{mn}=\s_{mn}{}^{\a\b}M_{\a\b},\quad {\rm
or}\quad M_{\a\b}=\frac{1}{2}M^-_{mn}\s_{mn\a\b}. \ee The above
equations are the precise statement of (\ref{sp2esu2}). 
So it may be more appropriate to say that the set of $SU(2)_R$
generators are related to the set of $Sp(2)$ generators via a
\emph{complex} linear nonsingular transformation.

One can use the first equation of (\ref{dltcdn1}) to verify that
$\s_{mn}{}^{\a\b}M_{\a\b}$ also obey the anti-duality condition
and satisfy the same commutation relations as $M^-_{mn}$ do. On
the other hand, we connect $M^+_{mn}$ with another set of
independent $SU(2)$ generators $M_{\dot\a\dot\b}$ by the equation
\e\label{2su211}
M^+_{mn}=\bar\s_{mn}{}^{\dot\a\dot\b}M_{\dot\a\dot\b}.
\ee In summary, we have \e\label{dcmof}
M_{mn}=M^-_{mn}+M^+_{mn}=\s_{mn}{}^{\a\b}M_{\a\b}+\bar\s_{mn}{}^{\dot\a\dot\b}M_{\dot\a\dot\b}.
\ee With the non-singular transformations (\ref{2su2}), we can
recast (\ref{cmrsp22n3}) into the following form:
\e\label{cmrsp22n3n} &&\{Q_{\bi\a},
Q_{\bj\b}\}=k^\prime(\ep_{\a\b}M_{\bi\bj}
+\d_{\bi\bj}M^-_{mn}\s_{mn\a\b}),\quad [M^-_{mn}, Q_{\bi\g}]=-\s_{mn\g}{}^{\a}Q_{\bi\a},\nonumber\\
&&[M_{\bi\bj},Q_{\bk\a}]=\d_{\bj\bk}Q_{\bi\a}-\d_{\bi\bk}Q_{\bj\a}.
\ee Here it is important to note that we can use either
(\ref{cmrsp22n3}) or (\ref{cmrsp22n3n}) to calculate the structure
constants $f_{ab\pa\pb}$ and $f_{\pa\pb\pc\pd}$. The final results
are the same. In section \ref{secGN4}, we will provide an
explanation on this.

Now it is straightforward to calculate the structure constants
$f_{ab\pa}{}^\pb$ by using Eq. (\ref{mbrcks2}): \e
[\{Q_a, Q_b\},Q_\pa]&=&[\{Q_{m\hi}, Q_{n\hj}\},Q_{\bi\a}]\nonumber\\
&=&[\omega_{\hi\hj}M^-_{mn},Q_{\bi\a}]\nonumber\\
&=&-k\omega_{\hi\hj}\s_{mn\a}{}^\b Q_{\bi\b}. \ee In the final
line we have used the second equation of (\ref{cmrsp22n3n}). Now
$f_{ab\pa\pb}$ can be read off from the above equation
immediately: \e\label{strct20}
f_{ab\pa\pb}=f_{m\hi,n\hj,\bi\a,\bj\b}=-k\omega_{\hi\hj}\d_{\bi\bj}\s_{mn\a\b}.
\ee
The structure constants $f_{\pa\pb ab}$ can be also read off
from $[\{Q_\pa, Q_\pb\}, Q_a]=f_{\pa\pb a}{}^bQ_b$. By requiring
that $f_{ab\pa\pb}=f_{\pa\pb ab}$, $k^\prime$ in (\ref{cmrsp22n3})
is determined to be $k^\prime=-\frac{1}{2}k$.

Since both $G$ and $G^\prime$ are ortho-symplectic algebras, the
structure constants $f_{abcd}$ and $f_{\pa\pb\pc\pd}$ must take
the same form as (\ref{strct2}); they are given by
\begin{eqnarray}\label{strct21}
f_{abcd}=f_{m\hat{i},n\hat{j},p\hat{k},q\hat{l}}
=k[(\delta_{mp}\delta_{nq}-\delta_{mq}\delta_{np})
\omega_{\hat{i}\hat{j}}\omega_{\hat{k}\hat{l}}
-\delta_{mn}\delta_{pq}(\omega_{\hat{i}\hat{k}}
\omega_{\hat{j}\hat{l}}+\omega_{\hat{i}\hat{l}}
\omega_{\hat{j}\hat{k}})],\\
f_{\pa\pb\pc\pd}=f_{\bi\a,\bj\b,\bk\g,\bl\d}
=-\frac{k}{2}[(\delta_{\bi\bk}\delta_{\bj\bl}-\delta_{\bi\bl}\delta_{\bj\bk})
\ep_{\a\b}\ep_{\g\d}
-\delta_{\bi\bj}\delta_{\bk\bl}(\ep_{\a\g}
\ep_{\b\d}+\ep_{\a\d}
\ep_{\b\g})].\label{strct22}
\end{eqnarray}
It is straightforward to verify that the structure constants
(\ref{strct20}), (\ref{strct21}) and (\ref{strct22}) satisfy the
symmetry conditions (\ref{symfs}) and the reality conditions
(\ref{rltcndtn}), and obey the fundamental identities (\ref{FI4}).
In verifying the FIs, we have used the identity
\begin{equation}
\s_{mp\a\b}\s_{pn\g\d}-\s_{np\a\b}\s_{pm\g\d}=\frac{1}{2}(\ep_{\a\g}\s_{mn\b\d}+
\ep_{\b\g}\s_{mn\a\d}+\ep_{\a\d}\s_{mn\b\g}+
\ep_{\b\d}\s_{mn\a\g}).
\end{equation}
Substituting (\ref{strct20}), (\ref{strct21}) and (\ref{strct22})
into (\ref{n4cov}), (\ref{LN4}) and (\ref{SUSY4}) gives the
$\CN=4$, $Sp(2N)\times SU(2)_R\times SU(2)_L\times SO(M)$ theory.
In this realization, the un-twisted multiplets are in the
bifundamental representation of $Sp(2N)\times SO(4)$, while the
twisted multiplets are in the bifundamental representation of
$Sp(2)\times SO(M)$.

We now would like to work out the explicit expression of the gauge
field $\tilde{A}_{\mu ab}$. Substituting the expressions of
$f_{cdab}$ and $f_{\pa\pb ab}$ (see (\ref{strct20}) and
(\ref{strct21})) into the second equation of (\ref{n4cov}), the
definition of $\tilde{A}_{\mu ab}$, we obtain
\begin{equation}\label{tutgauge}
\tilde{A}_{\mu ab}=(\tilde{A}_{\mu})_{m\hi,n\hj}
=k\omega_{\hi\hj}[A^{pq}_\mu (\t^+_{pq})_{mn}+(A^{pq}_\mu
-\frac{1}{2}A^{\prime\a\b}\s^{pq}{}_{\a\b})(\t^-_{pq})_{mn}]
+k\d_{mn}A^{\hk\hl}_{\mu}(\t_{\hk\hl})_{\hi\hj}.
\end{equation}
where we have used $A^{cd}_\mu=A^{p\hk,q\hl}_\mu$ and $A^{\pa\pb}_\mu=A^{\bi\a,\bj\b}_\mu$,
and  defined $A^{pq}_\mu\equiv \omega_{\hk\hl}A^{p\hk,q\hl}_\mu$ and
$A^{\prime\a\b}\equiv\d_{\bi\bj}A^{\bi\a,\bj\b}_\mu$.
The two sets of matrices $(\t^{\pm}_{pq})_{mn}$ are the vector
representations of the $SU(2)_{L,R}$ generators $M^{\pm}_{pq}$,
while $(\t_{\hk\hl})_{\hi\hj}$ are the fundamental representation
of the set of $Sp(2N)$ generators $M_{\hk\hl}$; they are given by
\e\label{rpoflr}
(\t^{\pm}_{pq})_{mn}=\frac{1}{2}(\d_{mp}\d_{nq}-\d_{np}\d_{mq}\pm
\vp_{mnpq}),\quad (\t_{\hk\hl})_{\hi\hj}=-(\omega_{\hat{i}\hat{k}}
\omega_{\hat{j}\hat{l}}+\omega_{\hat{i}\hat{l}}
\omega_{\hat{j}\hat{k}}). \ee If $f_{\pa\pb ab}$=0, the theory
becomes two copies of independent GW theories, and the gauge field
$\tilde{A}_{\mu ab}$ becomes \e\label{gwgauge} (\tilde{A}_{\mu
ab})_{f_{ab\pa\pb}=0}=k\omega_{\hi\hj}A^{pq}_\mu
[(\t^+_{pq})_{mn}+(\t^-_{pq})_{mn}]+k\d_{mn}A^{\hk\hl}_{\mu}(\t_{\hk\hl})_{\hi\hj},
\ee where the matrices \e\label{rpso4}
(\t^+_{pq})_{mn}+(\t^-_{pq})_{mn}=(\t_{pq})_{mn}=\delta_{mp}\delta_{nq}-\delta_{mq}\delta_{np}
\ee furnish the familiar vector representation of $SO(4)$. The RHS
of (\ref{gwgauge}) indicates that the Chern-Simons levels of
$SU(2)_R$ and $SU(2)_L$ gauge groups are equal and \emph{same} in
sign. Comparing (\ref{tutgauge}) and (\ref{gwgauge}), we see that
the structure constants $f_{ab\pa\pb}$ play a crucial role in
constructing the $\CN=4$ quiver gauge theory. Similarly, we obtain
the expression of the gauge field $\tilde{A}_{\mu\pa\pb}$ defined
in the last equation of (\ref{n4cov}): \e
\tilde{A}_{\mu\pa\pb}=-\frac{k}{2}[\d_{\bi\bj}(A^{\prime\g\d}_\mu
-A^{pq}_\mu\s_{pq}{}^{\g\d})(\t_{\g\d})_{\a\b}+\ep_{\a\b}A^{\prime\bk\bl}_\mu(\t_{\bk\bl})_{\bi\bj}].
\ee
Here $A^{\prime\bk\bl}_\mu\equiv \ep_{\g\d}A^{\bk\g,\bl\d}$;
$(\t_{\g\d})_{\a\b}$ and $(\t_{\bk\bl})_{\bi\bj}$, the fundamental
representations of $M_{\g\d}$ and $M_{\bk\bl}$, have similar
expressions as that of $(\t_{\hk\hl})_{\hi\hj}$ and
$(\t_{pq})_{mn}$ (see (\ref{rpoflr}) and (\ref{rpso4})),
respectively. Although we set $SU(2)_R=Sp(2)$ by Eq. (\ref{2su2}),
their representations are completely different: The set of
fermionic generators $Q_{m\hi}$ furnish a vector representation of
the $SU(2)_R$ generators $M_{mn}^-$,  while $Q_{\bi\g}$ furnish
the fundamental representation of the $Sp(2)$ generators
$M_{\a\b}$.

Also, If $f_{\pa\pb ab}$=0, the gauge field
$\tilde{A}_{\mu\pa\pb}$ becomes \e\label{gwgauge2}
(\tilde{A}_{\mu\pa\pb})_{f_{ab\pa\pb}=0}=
-\frac{k}{2}[\d_{\bi\bj}A^{\prime\g\d}_\mu
(\t_{\g\d})_{\a\b}+\ep_{\a\b}A^{\prime\bk\bl}_\mu(\t_{\bk\bl})_{\bi\bj}],
\ee which is just a gauge field of the GW theory.

Using the same technique, one can also pair $G=OSp(4|2N)$ with
other superalgebras, such as $OSp(3|2N_1)$. In summary, one can
pair $G=OSp(4|2N)$ with
\begin{equation}\label{ex1}
G^\prime=OSp(M|2),\quad OSp(3|2N_1),\quad PSU(2|2), \quad
G_3,\quad SU(2|N_2),\quad F(4),\quad {\rm or} \quad D(2|1,\alpha)
\end{equation}
by identifying $SU(2)_R$ or $SU(2)_L$ factor of $G$ with a simple
factor $SU(2)$ (or its isometries $Sp(2)$ and $SO(3)$) contained
in the bosonic part of $G^\prime$.

Finally, one can pair $G=D(2|1,\alpha)$ with \e\label{ex2}
G^\prime=OSp(3|N_1),\quad OSp(M|2),\quad PSU(2|2),\quad G_3,\quad
SU(2|N_2),\quad{\rm or}\quad F(4) \ee by identifying $SU(2)_R$ or
$SU(2)_L$ factor of $D(2|1,\alpha)$ with a simple factor $SU(2)$
(or its isometries $Sp(2)$ and $SO(3)$) of the bosonic part of
$G^\prime$.

The pairs of superalgebras $G$ and  $G^\prime$ in (\ref{ex1}) and
(\ref{ex2}) can be used to construct the four 3-brackets (\ref{own3brcks}) and (\ref{mbrcks}) and  the
four FIs (\ref{FI4}). The Lie algebras of the gauge groups are just the
bosonic parts of $G$ and  $G^\prime$, and corresponding
representations are determined by the fermionic generators of $G$
and  $G^\prime$.

We note that the bosonic part of $PSU(2|2)$ is $SO(4)$, we
therefore can identify it with the $SO(4)$ factor of the bosonic
part of $OSp(4|2N)$ or $D(2|1,\alpha)$. The special case of
$PSU(2|2)$ paring $OSp(4|2N)$ is interesting in its own right:
First, since the Nambu 3-algebra can be constructed in terms of
$PSU(2|2)$ \cite{Chen:pku1}, so the Nambu 3-algebra can be a sub
3-algebra of a symplectic 3-algebra; secondly, since the set of
$SO(4)$ generators of $OSp(4|2N)$ are given by
$(M^+_{mn}+M^-_{mn})=M_{mn}$ (see (\ref{dcmof})), while the set of
$SO(4)$ generators of $PSU(2|2)$ can be converted into
$(M^+_{mn}-M^-_{mn})=\tilde M_{mn}$ \cite{Chen:pku1}, so in
analogue with (\ref{2su2}), the two sets of $SO(4)$ generators can
be related to each other via the following duality equation: \e
(M_{mn})_{OSp(4|2N)}=\frac{1}{2}\vp_{mnpq}(\tilde
M_{pq})_{PSU(2|2)}. \ee

Notice also that we can identify the $SO(4)$ factor of the bosonic
part of $OSp(4|2N)$ with the $SO(4)$ factor of the bosonic part of
$D(2|1,\alpha)$, and the resulting gauge group is \emph{different}
from the one derived by letting that $OSp(4|2N)$ and
$D(2|1,\alpha)$ share only \emph{one} $SU(2)$ factor. The duality
equation for identifying two $SO(4)$ factors can be given by \e
\frac{1}{2}(\d_{mp}\d_{nq}-\d_{np}\d_{mq}+\b\vp_{mnpq})(M_{pq})_{OSp(4|2N)}=(\hat
M_{mn})_{D(2|1,\alpha)} \ee where $\b=2(1-\a)/(1+\a)$. Similarly,
the duality equation for identifying the $SO(4)$ factor of
$D(2|1,\alpha)$ with the $SO(4)$ factor of $PSU(2|2)$ can be given
by \e
\frac{1}{2}[\b(\d_{mp}\d_{nq}-\d_{np}\d_{mq})+\vp_{mnpq})](\tilde
M_{pq})_{PSU(2|2)}=(\hat M_{mn})_{D(2|1,\alpha)}. \ee We summarize
the three pairs as follows
\begin{equation}\label{idso4}
(G,G^\prime)=(PSU(2|2), OSp(4|2N)), (PSU(2|2), D(2|1,\alpha)), (OSp(4|2N), D(2|1,\alpha)).
\end{equation}
Every pair of superalgebras in the right hand side, whose bosonic
parts share the common factor $SO(4)$, can be used to construct
the four 3-brackets (\ref{own3brcks}) and (\ref{mbrcks}) and  the
four FIs (\ref{FI4}).

\subsection{$U(N_1)\times U(N_2)\times U(N_3)\times
U(N_4)$}\label{sec4un} Here we choose the superalgebras $G$ and
$G^\prime$ (see (\ref{slie2}) and (\ref{slie3})) as $U(N_1|N_2)$
and $U(N_3|N_4)$ respectively (the commutation relations of
$U(M|N)$ are given by Appendix \ref{cmrunum}). We identify the
$U(1)$ parts of the bosonic subalgebras of $U(N_1|N_2)$ and
$U(N_3|N_4)$ by introducing the following commutators between the
generators of $U(N_1|N_2)$ and the generators of $U(N_3|N_4)$:
\e\label{chrgdQ} &&[M_{\bu}{}^{\bv},
Q_{\bi}{}^{\hi}]=c_1\d_{\bu}{}^{\bv}Q_{\bi}{}^{\hi},\quad
[M_{\bu}{}^{\bv}, \bar Q_{\hi}{}^{\bi}]=-c_1\d_{\bu}{}^{\bv}\bar Q_{\hi}{}^{\bi},\nonumber\\
&&[M_{\pri}{}^{\pj}, Q_{\bi}{}^{\hi}]=c_2\d_{\pri}{}^{\pj}Q_{\bi}{}^{\hi},\quad
[M_{\pri}{}^{\pj}, \bar Q_{\hi}{}^{\bi}]=-c_2\d_{\pri}{}^{\pj}\bar Q_{\hi}{}^{\bi},\nonumber\\
&&[M_{\bi}{}^{\bj}, Q_{\bu}{}^\pri]=c_3\d_{\bi}{}^{\bj}Q_{\bu}{}^\pri,\quad
[M_{\bi}{}^{\bj}, \bar Q_{\pri}{}^{\bu}]=-c_3\d_{\bi}{}^{\bj}\bar Q_{\pri}{}^{\bu},
\nonumber\\
&&[M_{\hi}{}^{\hj},
Q_{\bu}{}^\pri]=c_4\d_{\hi}{}^{\hj}Q_{\bu}{}^\pri,\quad
[M_{\hi}{}^{\hj}, \bar Q_{\pri}{}^{\bu}]=-c_4\d_{\hi}{}^{\hj}\bar
Q_{\pri}{}^{\bu}, \ee where $Q_{\bi}{}^{\hi}$ are the fermionic
generators of $U(N_1|N_2)$, with $\bi=1,\ldots, N_1$ fundamental
indices of $U(N_1)$ and $\hi=1,\ldots, N_2$ anti-fundamental
indices of $U(N_2)$; $M_{\bi}{}^{\bj}$ and $M_{\hi}{}^{\hj}$ are
the bosonic generators of $U(N_1|N_2)$; $Q_{\bu}{}^\pri$ are the
fermionic generators of $U(N_3|N_4)$, with $\bu=1,\ldots, N_3$
fundamental indices of $U(N_3)$ and $\pri=1,\ldots, N_4$
anti-fundamental indices of $U(N_4)$; $M_{\bu}{}^{\bv}$ and
$M_{\pri}{}^{\pj}$ are the bosonic generators of $U(N_3|N_4)$;
$c_i$ ($i=1,\ldots, 4$) are arbitrary constants.

Let us now examine the physical meaning of the first commutator of
(\ref{chrgdQ}). Writing the set of $U(N_3)$ generators
$M_{\bu}{}^{\bv}$ as \e
(M_{\bu}{}^{\bv})_{U(N_3)}=(M_{\bu}{}^{\bv}-\frac{1}{N_3}\d_{\bu}{}^{\bv}M_{\bw}{}^{\bw})_{SU(N_3)}
+(\frac{1}{N_3}\d_{\bu}{}^{\bv}M_{\bw}{}^{\bw})_{U(1)}, \ee and
using the first commutator of (\ref{chrgdQ}), we find that
$Q_{\bi}{}^{\hi}$ commutes with the set of $SU(N_3)$ generators,
while has a nontrivial commutator with the $U(1)$ generator of
$U(N_3)$, i.e. \e\label{chrgdQ2}
[M_{\bu}{}^{\bv}-\frac{1}{N_3}\d_{\bu}{}^{\bv}M_{\bw}{}^{\bw},
Q_{\bi}{}^{\hi}]=0 ,\quad
[\frac{1}{N_3}\d_{\bu}{}^{\bv}M_{\bw}{}^{\bw},
Q_{\bi}{}^{\hi}]=c_1\d_{\bu}{}^{\bv}Q_{\bi}{}^{\hi}. \ee So the
first commutator of (\ref{chrgdQ}) means that the set of fermionic
generators of $U(N_1|N_2)$ $Q_{\bi}{}^{\hi}$ are charged by the
$U(1)$ part of $U(N_3)$ of the bosonic part of $U(N_3|N_4)$; also,
it means that $Q_{\bi}{}^{\hi}$ furnish a nontrivial
representation of the $U(1)$ part of $U(N_3)$ of the bosonic part
of $U(N_3|N_4)$. The other commutators in (\ref{chrgdQ}) have a
similar interpretation.

On the other hand, let us consider the following commutator of
$U(N_1|N_2)$: $[M_{\bj}{}^{\bk},
Q_{\bi}{}^{\hi}]=\d_{\bi}{}^{\bk}Q_{\bj}{}^{\hi}$. Contracting on
$\bj$ and $\bk$ gives \e\label{chrgdQ3}
[\frac{1}{N_1}M_{\bj}{}^{\bj},
Q_{\bi}{}^{\hi}]=\frac{1}{N_1}Q_{\bi}{}^{\hi}. \ee Namely,
$Q_{\bi}{}^{\hi}$ are also charged by the $U(1)$ part of $U(N_1)$
of the bosonic part of $U(N_1|N_2)$, or $Q_{\bi}{}^{\hi}$  furnish
a nontrivial representation of the $U(1)$ part of $U(N_1)$ of the
bosonic part of $U(N_1|N_2)$. Similarly, $Q_{\bi}{}^{\hi}$ furnish
a nontrivial representation of the $U(1)$ part of $U(N_2)$ of the
bosonic part of $U(N_1|N_2)$.

So, after identifying the $U(1)$ parts of the bosonic subalgebras
of $U(N_1|N_2)$ and $U(N_3|N_4)$, both $Q_{\bi}{}^{\hi}$ and
$Q_{\bu}{}^\pri$ have nontrivial commutators with \emph{all}
$U(1)$ generators of $U(N_1|N_2)$ \emph{and} $U(N_3|N_4)$.

We are now ready to calculate the structure constants of the
double graded commutators. We define \e
&&Q_a=\begin{pmatrix} \bar Q_{\bi}{}^{\hi}\\-Q_{\hi}{}^{\bi}\end{pmatrix}=\bar Q_{\bi}{}^{\hi}\d_{1\lambda}-Q_{\hi}{}^{\bi}\d_{2\lambda},\\
&& Q_\pa=\begin{pmatrix} \bar
Q_\pri{}^{\bu}\\-Q_{\bu}{}^\pri\end{pmatrix}=\bar
Q_\pri{}^{\bu}\d_{1\a}-Q_{\bu}{}^\pri\d_{2\a}, \ee where
$\d_{1\lambda}=(1,0)^{{\rm T}}$ and $\d_{2\lambda}=(0,1)^{{\rm
T}}$ are ``spin up" spinor (\emph{not} a spacetime spinor) and
``spin down" spinor, respectively. Similarly, $\d_{1\a}$ and
$\d_{2\a}$ are another independent pair of spinors. Using
$[\{Q_a,Q_b\},Q_\pa]=f_{ab\pa}{}^\pb Q_\pb$, and the commutation
relations of $U(N_1|N_2)$ and (\ref{chrgdQ}), we obtain
\begin{equation}\label{strct88}
f_{ab\pa\pb}=-k(c_1+c_2)(\d_{\hi}{}^{\hj}\d_{\bj}{}^{\bi}\d_{2\lambda}\d_{1\xi}+
\d_{\hj}{}^{\hi}\d_{\bi}{}^{\bj}\d_{1\lambda}\d_{2\xi})
(\d_{\bv}{}^{\bu}\d_{\pri}{}^{\pj}\d_{1\a}\d_{2\b}+
\d_{\bu}{}^{\bv}\d_{\pj}{}^{\pri}\d_{2\a}\d_{1\b}).
\end{equation}
Using $[\{Q_\pa,Q_\pb\},Q_a]=f_{\pa\pb a}{}^b Q_b$ to calculate
$f_{\pa\pb ab}$ gives the same result as that of (\ref{strct88})
except that $(c_1+c_2)$ gets replaced by $(c_3+c_4)$. In order for
that $f_{\pa\pb ab}=f_{ab\pa\pb }$, we must set \e
(c_1+c_2)=(c_3+c_4). \ee It can be seen that this $\CN=4$ theory
contains \emph{three free parameters}. The second FI and third FI
of (\ref{FI4}) do not impose any further constraint on $(c_1+c_2)$
or $(c_3+c_4)$ , since for example the summation of the first two
terms of the second FI of (\ref{FI4}) vanishes due to the abelian
nature of $U(1)$.

Using $[\{Q_\pa,Q_\pb\},Q_\pc]=f_{\pa\pb\pc}{}^\pd Q_\pd$ and the commutation relations of $U(N_3|N_4)$, one can calculate the structure constants $f_{\pa\pb\pc\pd}$ straightforwardly; they are given by
\begin{eqnarray}\label{strct87}
f_{\pa\pb\pc\pd}
&=&f_{\bu}{}^\pri{}_{\bw}{}^\pk,{}_\pl{}^{\bt}{}_\pj{}^{\bv}\delta_{2\alpha}\delta_{1\beta}\delta_{2\gamma}\delta_{1\delta}
+f_{\bu}{}^\pri{}_{\bt}{}^\pl,{}_\pk{}^{\bw}{}_\pj{}^{\bv}\delta_{2\alpha}\delta_{1\beta}\delta_{1\gamma}\delta_{2\delta}
\nonumber\\&&+f_{\bv}{}^\pj{}_{\bw}{}^\pk,{}_\pl{}^{\bt}{}_\pri{}^{\bu}\delta_{1\alpha}\delta_{2\beta}\delta_{2\gamma}\delta_{1\delta}
+f_{\bv}{}^\pj{}_{\bt}{}^\pl,{}_\pk{}^{\bw}{}_\pri{}^{\bu}\delta_{1\alpha}\delta_{2\beta}\delta_{1\gamma}\delta_{2\delta},
\end{eqnarray}
where \e\label{strumun}
f_{\bu}{}^\pri{}_{\bv}{}^\pj,{}_\pk{}^{\bw}{}_\pl{}^{\bt} \equiv
k(\d_\pk{}^\pri\d_\pl{}^\pj\d_{\bv}{}^{\bw}\d_{\bu}{}^{\bt}
-\d_\pl{}^\pri\d_\pk{}^\pj\d_{\bu}{}^{\bw}\d_{\bv}{}^{\bt}). \ee
Notice that (\ref{strumun}) are precisely the structure constants
introduced first by Lambert and Bagger \cite{Bagger08:3Alg} to
construct an $\CN=6$ theory with $U(N_3)\times U(N_4)$ gauge
group. The relation between (\ref{strumun}) and (\ref{strct87})
was first derived in Ref. \cite{Chen2}. Since both $U(N_1|N_2)$
and $U(N_3|N_4)$ are unitary superalgebras, the structure
constants $f_{abcd}$ have a similar expression as that of
(\ref{strct87}); they are given by
\begin{eqnarray}\label{strct86}
f_{abcd}
&=&f_{\bi}{}^{\hi}{}_{\bk}{}^{\hk},{}_{\hl}{}^{\bl}{}_{\hj}{}^{\bj}\delta_{2\lambda}\delta_{1\xi}\delta_{2\rho}\delta_{1\sigma}
+f_{\bi}{}^{\hi}{}_{\bl}{}^{\hl},{}_{\hk}{}^{\bk}{}_{\hj}{}^{\bj}\delta_{2\lambda}\delta_{1\xi}\delta_{1\rho}\delta_{2\sigma}
\nonumber\\&&+f_{\bj}{}^{\hj}{}_{\bk}{}^{\hk},{}_{\hl}{}^{\bl}{}_{\hi}{}^{\bi}\delta_{1\lambda}\delta_{2\xi}\delta_{2\rho}\delta_{1\sigma}
+f_{\bj}{}^{\hj}{}_{\bl}{}^{\hl},{}_{\hk}{}^{\bk}{}_{\hi}{}^{\bi}\delta_{1\lambda}\delta_{2\xi}\delta_{1\rho}\delta_{2\sigma},
\end{eqnarray}
where \e\label{strumun2}
f_{\bi}{}^{\hi}{}_{\bj}{}^{\hj},{}_{\hk}{}^{\bk}{}_{\hl}{}^{\bl}
\equiv
k(\d_{\hk}{}^{\hi}\d_{\hl}{}^{\hj}\d_{\bj}{}^{\bk}\d_{\bi}{}^{\bl}
-\d_{\hl}{}^{\hi}\d_{\hk}{}^{\hj}\d_{\bi}{}^{\bk}\d_{\bj}{}^{\bl}).
\ee We have verified that the structure constants (\ref{strct88}),
(\ref{strct87}) and (\ref{strct86}) obey the fundamental
identities (\ref{FI4}), and satisfy the desired symmetry and
reality conditons. Substituting them into (\ref{n4cov}),
(\ref{LN4}) and (\ref{SUSY4}) gives the $\CN=4$, $U(N_1)\times
U(N_2)\times U(N_3)\times U(N_4)$ theory. In this realization, the
un-twisted multiplets are in the bifundamental representation of
$U(N_1)\times U(N_2)$, while the twisted multiplets are in the
bifundamental representation of $U(N_3)\times U(N_4)$. However,
the un-twisted multiplets couple the twisted multiplets
nontrivially via the structure constants $f_{ab\pa\pb}$. In the
special case of $c_1+c_2=0$, the structure constants
$f_{ab\pa\pb}$ vanish identically. As a result, the action
(\ref{LN4}) becomes two uncoupled GW theories.

\subsection{$OSp(N_2)\times U(1)\times U(N_3)\times
U(N_4)$}\label{secospu} In this subsection we choose the
superalgebras $G$ and $G^\prime$ (see (\ref{slie2}) and
(\ref{slie3})) as $OSp(2|2N_2)$ and $U(N_3|N_4)$ respectively.
(The commutation relations of $OSp(M|2N)$ and $U(M|N)$ are given
by Appendix \ref{cmrsp22n} and \ref{cmrunum} respectively.) In
analogy to Section \ref{sec4un}, we identify the $U(1)$ parts of
the bosonic subalgebras of $OSp(2|2N_2)$ and $U(N_3|N_4)$ by
introducing the following commutators between their generators:
\e\label{chrgdQ10} &&[M_{\bu}{}^{\bv},
Q_{\bi\hi}]=c_1\d_{\bu}{}^{\bv}\ep_{\bi\bj}Q_{\bj\hi},\quad
[M_{\pri}{}^{\pj}, Q_{\bi\hi}]=c_2\d_{\pri}{}^{\pj}\ep_{\bi\bj}Q_{\bj\hi},\nonumber\\
&&[M_{\bi\bj},
Q_{\bu}{}^\pri]=-c_3\ep_{\bi\bj}Q_{\bu}{}^\pri,\quad [M_{\bi\bj},
\bar Q_{\pri}{}^{\bu}]=c_3\ep_{\bi\bj}\bar Q_{\pri}{}^{\bu}, \ee
where $Q_{\bi\hi}$ are the set of fermionic generators of
$OSp(2|2N_2)$, with $\bi=1,2$ fundamental indices of $SO(2)$ and
$\hi=1,\ldots, N_2$ fundamental indices of $Sp(2N_2)$;
$M_{\bi\bj}$ and $M_{\hi\hj}$ are the bosonic generators of
$OSp(2|2N_2)$; $Q_{\bu}{}^\pri$ are the set of fermionic
generators of $U(N_3|N_4)$, with $\bu=1,\ldots, N_3$ fundamental
indices of $U(N_3)$ and $\pri=1,\ldots, N_4$ anti-fundamental
indices of $U(N_4)$; $M_{\bu}{}^{\bv}$ and $M_{\pri}{}^{\pj}$ are
the bosonic generators of $U(N_3|N_4)$; $c_i$ ($i=1,\ldots, 3$)
are arbitrary constants; $\ep_{\bi\bj}=-\ep_{\bj\bi}$ and
$\ep_{\bar 1\bar 2}=1$. The commutators in (\ref{chrgdQ10}) have
similar interpretations as that of section \ref{sec4un}.

To calculate the structure constants $f_{\pa\pb ab}$, we define \e
Q_a=Q_{\bi\hi}, \quad Q_\pa=\begin{pmatrix} \bar Q_\pri{}^{\bu}\\
-Q_{\bu}{}^\pri\end{pmatrix}=\bar
Q_\pri{}^{\bu}\d_{1\a}-Q_{\bu}{}^\pri\d_{2\a}, \ee where
$\d_{1\a}=(1,0)^{{\rm T}}$ and $\d_{2\a}=(0,1)^{{\rm T}}$ are
``spin up" spinor (\emph{not} a spacetime spinor) and ``spin down"
spinor, respectively. Using $[\{Q_\pa,Q_\pb\},Q_a]=f_{\pa\pb
a}{}^b Q_b$ , and the commutation relations of $U(N_2|N_3)$ and
(\ref{chrgdQ10}), we obtain
\begin{equation}\label{strct98}
f_{ab\pa\pb}=k(c_1+c_2)
(\d_{\bv}{}^{\bu}\d_{\pri}{}^{\pj}\d_{1\a}\d_{2\b}+
\d_{\bu}{}^{\bv}\d_{\pj}{}^{\pri}\d_{2\a}\d_{1\b})\ep_{\bi\bj}\omega_{\hi\hj}.
\end{equation}
One can also use $[\{Q_a,Q_b\},Q_\pa]=f_{ab\pa}{}^\pb Q_\pb$ to
calculate $f_{ ab\pa\pb}$. The final expression is of course the
same as that of (\ref{strct98}) after we set \e c_1+c_2=c_3. \ee
So this $\CN=4$ theory contains \emph{two free parameters}. By the
same reason as that of Section \ref{sec4un}, the second FI and
third FI of (\ref{FI4}) impose no constraint on $(c_1+c_2)$ or
$c_3$. The structure constants $f_{abcd}$ and $f_{\pa\pb\pc\pd}$
are the same as (\ref{strct2}) and (\ref{strct87}) respectively,
except for that here the index $a=1,\ldots, N_2$. We have also
verified that all the structure constants obey fundamental
identities (\ref{FI4}) and satisfy the desired reality and
symmetry conditons. Notice that section \ref{secospunu1} is a
special case of this subsection. Actually, if we set $N_4=1$,
$N_2=N$ and $c_3=1$, the structure constants $f_{abcd}$,
$f_{ab\pa\pb}$ and $f_{\pa\pb\pc\pd}$ are exactly the same as
(\ref{strct2}), (\ref{strct1}) and  (\ref{strct3}), respectively.

Substituting the structure constants of this subsection
$f_{abcd}$, $f_{ab\pa\pb}$ and $f_{\pa\pb\pc\pd}$ into
(\ref{n4cov}), (\ref{LN4}) and (\ref{SUSY4}) gives the $\CN=4$,
$Sp(2N_2)\times U(1)\times U(N_3)\times U(N_4)$ theory. The
un-twisted multiplets are in the bifundamental representation of
$Sp(2N_2)\times U(1)$, while the twisted multiplets are in the
bifundamental representation of $U(N_3)\times U(N_4)$.

\section{General $\CN=4$ Theories in Terms of Lie
Algebras}\label{secGN4} In this section, we will derive the
general $\CN=4$ theories in terms of Lie algebras from their
3-algebra counterparts. The key point is that we observe that the
structure constants of 3-algebras can be expressed in terms of
tensor products on the superalgebras $G$ and $G^\prime$. Eqs
(\ref{tensorprd1}) are two examples.

Recall that the bosonic parts of the two superalgebras (\ref{slie2}) and
(\ref{slie3}) share at least one simple factor or $U(1)$ factor,
and we have decomposed their bosonic generators $M^u$ and $M^\pu$ into
$M^u=(M^\a,M^g)$ \footnote{Here $\a$ is \emph{not} a spacetime
index. We hope this will not cause any confusion.} and
$M^\pu=(M^{\a^\prime},\tilde M^g)$, respectively. Here $\tilde
M^g=T^g{}_hM^h$ are the generators of the common part, with
$T^g{}_h$ a set of \emph{complex} non-singular linear
transformation matrices. If $T^g{}_h$ are \emph{real} and positive
difinite, the two Lie algebras spanned by $M^g$ and $\tilde M^g$
are equivalent. In particular, if $T^g{}_h=\d^g{}_h$, we have
$\tilde M^g=M^g$. The independent commutation relations of the
bosonic parts of the two superalgebras (\ref{slie2}) and
(\ref{slie3}) are the following:
\begin{equation}\label{3kndcmm}
[M^\a,M^\b]=f^{\a\b}{}_\g M^\g,\quad [M^f,M^g]=f^{fg}{}_hM^h,\quad
[M^{\a^\prime},M^{\b^\prime}]=f^{\a^\prime\b^\prime}{}_{\g^\prime}
M^{\g^\prime}.
\end{equation}
Of course, we also have \e\label{nonindpnteq} [\tilde M^f,\tilde
M^g]=\tilde f^{fg}{}_h\tilde M^h. \ee However, since this equation
can be obtained by transforming the second equation of
(\ref{3kndcmm}) by using $\tilde M^g=T^g{}_hM^h$, we do not
consider it as an independent equation. Accordingly, we decompose
the structure constants into
\begin{equation}
f^{uv}{}_w=(f^{\a\b}{}_\g,f^{fg}{}_h),\quad f^{\pu\pv}{}_\pw=(f^{\a^\prime\b^\prime}{}_{\g^\prime},\tilde f^{fg}{}_h).
\end{equation}
Now the superalgebra (\ref{slie2}) is decomposed into
\e\label{slied1}
&&[M^\a,M^\b]=f^{\a\b}{}_\g M^\g,\quad [M^f,M^g]=f^{fg}{}_hM^h,\nonumber\\
&&[M^\a, Q_a]=-\t^\a_{ab}\omega^{bc}Q_c,\quad [M^g, Q_a]=-\t^g_{ab}\omega^{bc}Q_c,\nonumber\\
&& \{Q_a,Q_b\}=\t^\a_{ab}k_{\a\b}M^\b
+\t^g_{ab}k_{gh}M^h.
\ee
Similarly, the superalgebra (\ref{slie3}) can be written as
\e\label{slied3}
&&[M^\gap,M^\gbp]=f^{\gap\gbp}{}_\ggp M^\ggp,\quad [\tilde M^f,\tilde M^g]=\tilde f^{fg}{}_h\tilde M^h,\nonumber\\
&&[M^\gap, Q_{\pa}]=-\t^\gap_{\pa\pb}\omega^{\pb\pc}Q_{\pc},\quad [\tilde M^g, Q_{\pa}]=-\tilde \t^g_{\pa\pb}\omega^{\pb\pc}Q_{\pc},\nonumber\\
&& \{Q_{\pa},Q_{\pb}\}=\t^\gap_{\pa\pb}k_{\gap\gbp}M^\gbp
+\tilde\t^g_{\pa\pb}\tilde k_{gh}\tilde M^h.
\ee
Using the non-singular transformation $\tilde M^g=T^g{}_hM^h$, we are able to
recast ($\ref{slied3}$) in the form
\e\label{slied2}
&&[M^\gap,M^\gbp]=f^{\gap\gbp}{}_\ggp M^\ggp,\quad [M^f,M^g]=f^{fg}{}_hM^h,\nonumber\\
&&[M^\gap, Q_{\pa}]=-\t^\gap_{\pa\pb}\omega^{\pb\pc}Q_{\pc},\quad [ M^g, Q_{\pa}]=- \t^g_{\pa\pb}\omega^{\pb\pc}Q_{\pc},\nonumber\\
&& \{Q_{\pa},Q_{\pb}\}=\t^\gap_{\pa\pb}k_{\gap\gbp}M^\gbp
+\t^g_{\pa\pb}k_{gh}M^h,
\ee
where
\begin{equation}\label{nonsigtran}
\t^g_{\pa\pb}=(T^{-1})^g{}_h\tilde\t^h_{\pa\pb},\quad k_{gh}=T^f{}_gT^i{}_h\tilde k_{fi},
\quad f^{fg}{}_h=(T^{-1})^f{}_{i}(T^{-1})^g{}_{j}T^k{}_h\tilde f^{ij}{}_k.
\end{equation}
Since $T^g{}_h$ are
generally \emph{complex} nonsingular matrices, the
Lie algebra defined by the second equation of (\ref{slied2}) is
generally \emph{not} equivalent to the one defined by the second
equation of (\ref{slied3}). However, the $\CN=4$ theories will not
be modified if we use (\ref{slied2}) to construct the 3-brackets.
In fact, since we have not transformed the set of fermionic
generators $Q_\pa$ in (\ref{slied3}), the double graded brackets
$[\{Q_\pa,Q_\pb\}, Q_\pc]$ and $[\{Q_a,Q_b\}, Q_\pa]$ will be the
same no matter we use (\ref{slied3}) or (\ref{slied2}) to
construct them. As a result, the structure constants
$f_{\pa\pb\pc\pd}$ and $f_{\pa\pb ab}$ used to construct the
$\CN=4$ theories will also remain the same. For example, if we use
(\ref{slied3}) and (\ref{slied2}) to calculate $[\{Q_\pa,Q_\pb\},
Q_\pc]=f_{\pa\pb\pc}{}^\pd Q_\pd$ respectively, we obtain
\begin{equation}
f_{\pa\pb\pc\pd}=\t^\gap_{\pa\pb}k_{\gap\gbp}\t^\gbp_{\pc\pd}
+\tilde\t^g_{\pa\pb}\tilde k_{gh}\tilde\t^g_{\pc\pd}\quad{\rm and}\quad
f_{\pa\pb\pc\pd}=\t^\gap_{\pa\pb}k_{\gap\gbp}\t^\gbp_{\pc\pd}
+\t^g_{\pa\pb}k_{gh}\t^g_{\pc\pd}
\end{equation}
respectively. But it is not difficult to prove that
$\tilde\t^g_{\pa\pb}\tilde
k_{gh}\tilde\t^g_{\pc\pd}=\t^g_{\pa\pb}k_{gh}\t^g_{\pc\pd}$ by
using the first two equations of (\ref{nonsigtran}). So they are
indeed the same.

To simplify the expressions of (\ref{slied1}) and (\ref{slied2}),
we define \footnote{Here $m$, $n$ and $p$ are \emph{not} $SO(4)$
fundamental indices of Section \ref{secosp4n}. We hope this will
not cause any confusion.}
\begin{eqnarray}\label{drcsm1}
&&M^m=(M^\a,M^g,M^{\a^\prime}),\\
&&C^{mn}{}_p=(f^{\a\b}{}_\g,
f^{fg}{}_h,f^{\a^\prime\b^\prime}{}_{\g^\prime}),\\
&&k_{mn}=(k_{\a\b},k_{gh},k_{\a^\prime\b^\prime}).\label{drcsm3}
\end{eqnarray}
We now put the superalgebras (\ref{slied1}) \emph{and}
(\ref{slied2}) together: \e\label{Fsdspalg2}
&&[M^m,M^n]=C^{mn}{}_p M^p,\quad [M^m,
Q_a]=-\t^m_{ab}\omega^{bc}Q_c,\quad [M^m,
Q_\pa]=-\t^m_{\pa\pb}\omega^{\pb\pc}Q_\pc,\nonumber\\
&& \{Q_{a},Q_{b}\}=\t^m_{ab}k_{mn}M^n, \quad \{Q_{\pa},Q_{\pb}\}=\t^m_{\pa\pb}k_{mn}M^n,
\ee
where we have used the equations $\t^{\a^\prime}_{ab}=\t^\a_{\pa\pb}=0$ implied by (\ref{cmmt}).

Notice that (\ref{Fsdspalg2}) is merely a compact version of
(\ref{slied1}) \emph{and} (\ref{slied2}); in particular, it is
\emph{not} necessarily a closed superalgebra due to the set of
common generators $M^g$. In fact, because of $M^g$, the double
graded commutator $[\{Q_a, Q_b\},Q_\pc]\neq0$, i.e.
$f_{ab\pc}{}^\pd\neq0$ (see the proof in the paragraph containing
(\ref{cmmfct})). On the other hand, if (\ref{Fsdspalg2}) is a
closed superalgebra, then the $Q_aQ_bQ_\pc$ Jacobi identity must
be obeyed (see (\ref{JI1})), implying that $\{Q_a,Q_\pc\}\neq0$. Therefore the
anticommutator (\ref{mtm})
\e\label{ntrvanticm}
\{Q_a,Q_\pc\}=t^{\tilde{u}}_{a\pc}M_{\tilde{u}}\nonumber
\ee
must be nontrivial,
in the sense that $t^{\tilde{u}}_{a\pc}\neq0$. However,  the set of bosonic generators $M_{\tilde{u}}$ are \emph{not} included in $M^m$ (see
(\ref{drcsm1})), and (\ref{mtm}) is \emph{not} contained in
(\ref{Fsdspalg2}). So generally speaking, (\ref{Fsdspalg2}) is not a
closed superalgebra. But if we introduce $M_{\tilde{u}}$,
(\ref{mtm}) and some other proper commutation relations
into (\ref{Fsdspalg2}), it is possible to ``fuse" the two
superalgebras (\ref{slied1}) \emph{and} (\ref{slied2}) into a
single closed superalgebra \cite{ChenWu5}.

With these notations, if we construct the 3-brackets (\ref{4dgdcm}) and (\ref{mbrcks20}) by using
(\ref{Fsdspalg2}), the structure constants (see (\ref{tensorprd1}) and (\ref{tensorprd2})) for take the forms
\begin{equation}\label{n4ten} f_{abcd}=k_{mn}\t^m_{ab}\t^n_{cd},\quad
f_{\pa\pb\pc\pd}= k_{mn}\t^{m}_{\pa\pb}\t^{n}_{\pc\pd},\quad
f_{ab\pc\pd}=f_{\pc\pd ab}= k_{mn}\t^m_{ab}\t^{n}_{\pc\pd}.
\end{equation}
With the above equations, the four sets of FIs (\ref{FI4}) can be converted into
\e\label{C4}
&&(k_{np}k_{qm}C^{pm}{}_s+k_{qm}k_{sp}C^{pm}{}_n)\t^{na}{}_b\t^{qc}{}_d\t^{se}{}_f=0,\nonumber\\
&&(k_{np}k_{qm}C^{pm}{}_s+k_{qm}k_{sp}C^{pm}{}_n)\t^{na}{}_b\t^{q\pc}{}_\pd\t^{se}{}_f=0,
\nonumber\\
&&(k_{np}k_{qm}C^{pm}{}_s+k_{qm}k_{sp}C^{pm}{}_n)\t^{n\pa}{}_\pb\t^{q\pc}{}_\pd\t^{se}{}_f=0,
\nonumber\\
&&(k_{np}k_{qm}C^{pm}{}_s+k_{qm}k_{sp}C^{pm}{}_n)\t^{n\pa}{}_\pb\t^{q\pc}{}_\pd\t^{s\pe}{}_\prf=0.
\ee
They are simply obeyed due to the fact that
\begin{equation}
k_{np}k_{qm}C^{pm}{}_s+k_{qm}k_{sp}C^{pm}{}_n=0.
\end{equation}

With (\ref{n4ten}), the gauge fields (see (\ref{n4cov})) of the
$\CN=4$ theories become
\begin{eqnarray}\label{n4gauge}
&\tilde
A_\mu{}^c{}_d=A_\mu^{ab}f_{ab}{}^c{}_d+A_\mu^{\pa\pb}f_{\pa\pb}{}^c{}_d
=(A_\mu^{ab}\t^m_{ab}+A_\mu^{\pa\pb}\t^{m}_{\pa\pb})k_{mn}\t^{nc}{}_d
\equiv A_\mu^mk_{mn}\t^{nc}{}_d,\\
&\tilde A_\mu{}^\pc{}_\pd
=A_\mu^{ab}f_{ab}{}^\pc{}_\pd+A_\mu^{\pa\pb}f_{\pa\pb}{}^\pc{}_\pd
=(A_\mu^{ab}\t^m_{ab}+A_\mu^{\pa\pb}\t^m_{\pa\pb})k_{mn}\t^{n\pc}{}_\pd
\equiv A_\mu^mk_{mn}\t^{n\pc}{}_\pd\nonumber.
\end{eqnarray}
Following Ref. \cite{GaWi}, we define the `momentum map' and
`current ' operators as follows
\e \mu^m_{AB}\equiv \t^m_{ab}Z^a_AZ^b_B, \quad
j^m_{A\DB}\equiv\t^m_{ab}Z^a_A\p^b_\DB,\quad \mu^{\prime
m}_{\DA\DB}\equiv \t^m_{\pa\pb}Z^\pa_\DA Z^\pb_\DB, \quad j^{\prime
m}_{\DA B}\equiv\t^m_{\pa\pb}Z^\pa_\DA\p^\pb_B.\ee
Substituting (\ref{n4ten}) and (\ref{n4gauge}) into the $\CN=4$
Lagrangian (\ref{LN4}) gives
\e\label{2LN4}
\CL&=&\frac{1}{2}\epsilon^{\mu\nu\lambda}(k_{mn}A_\mu^m\partial_\nu
A_\lambda^n+\frac{1}{3}\tilde C_{mnp}A_\mu^mA_\nu^nA_\lambda^p)\nonumber\\
&&+\frac{1}{2}(-D_\mu\bar{Z}^A_aD^\mu Z^a_A-D_\mu\bar{Z}^\DA_\pa
D^\mu Z^\pa_\DA+i\bp^\DA_a\g^\mu D_\mu\p^a_\DA+i\bp^A_\pa\g^\mu
D_\mu\p^\pa_A)\nonumber\\&&-\frac{i}{2}k_{mn}(j^m_{A\DB}j^{nA\DB}
+j^{\prime m}_{\DA B}j^{\prime n\DA B}-4j^m_{A\DB}j^{\prime n\DB
A})\nonumber\\&&+\frac{i}{2}k_{mn}(\mu^m_{AB}\t^n_{\pa\pb}\p^{A\pa}\p^{B\pb}
+\mu^{\prime m}_{\DA\DB}\t^n_{ab}\p^{\DA a}\p^{\DB b})\\&&
-\frac{1}{24}\tilde C_{mnp}(\mu^{mA}{}_B\mu^{nB}{}_C\mu^{pC}{}_A
+\mu^{\prime m\DA}{}_\DB\mu^{\prime n\DB}{}_\DC\mu^{\prime
p\DC}{}_\DA)\nonumber\\&&+\frac{1}{4}k_{mp}k_{ns}((\t^m\t^n)_{ab}Z^{Aa}Z^b_A\mu^{\prime
p\DB}{}_\DC\mu^{\prime
s\DC}{}_\DB+(\t^m\t^n)_{\pa\pb}Z^{\DA\pa}Z^\pb_\DA\mu^{ pB}{}_C\mu^{
sC}{}_B)\nonumber, \ee
where $\tilde C_{mnp}=k_{ms}k_{nq}C^{sq}{}_p$. Substituting
(\ref{n4ten}) and (\ref{n4gauge}) into the $\CN=4$ supersymmetry
transformations (\ref{SUSY4}) gives
\e \label{2SUSY4}&&\delta Z^a_A=i\ep_A{}^\DA\p^a_\DA,\nonumber\\
&&\delta Z^\pa_\DA=i\ep^\dag_\DA{}^A\p^\pa_A,\nonumber\\
&&\delta\p^\pa_A=-\g^\mu D_\mu
Z^\pa_\DB\ep_A{}^\DB-\frac{1}{3}k_{mn}\t^{m\pa}{}_{\pb}Z^\pb_\DB\mu^{\prime
n\DB}{}_\DC\ep_A{}^\DC+k_{mn}\t^{m\pa}{}_\pb Z^\pb_\DA\mu^{nB}{}_A\ep_B{}^\DA, \nonumber\\
&&\delta\p^a_\DA=-\g^\mu D_\mu
Z^a_B\ep^\dag_\DA{}^B-\frac{1}{3}k_{mn}\t^{ma}{}_{b}Z^b_B\mu^{
nB}{}_C\ep^\dag_\DA{}^C+k_{mn}\t^{ma}{}_b Z^b_A\mu^{\prime n\DB}{}_\DA\ep^\dag_\DB{}^A,\nonumber\\
&&\delta A_\mu^m=i\ep^{A\DB}\g_\mu
j^m_{A\DB}+i\ep^{\dag\DA B}\g_\mu j^{\prime m}_{\DA B}.
\ee
Here the parameter $\ep_A{}^\DB$ obeys the reality condition
(\ref{n4para}). The $\CN=4$ Lagrangian (\ref{2LN4}) and
supersymmetry transformation law (\ref{2SUSY4}) are in agreement
with those constructed directly in terms of ordinary Lie algebra
\cite{HosomichiJD}.

If both the twisted and untwisted multiplets take values in the \emph{same} symplectic 3-algebra spanned by $T_a$, i.e. $\Phi_A=\Phi^a_A$ and $\Phi_{\dot A}=\Phi^a_{\dot A}$, we need only one superalgebra $G$ to construct the 3-algebra spanned by $T_a$. It follows that both $\Phi_A$ and $\Phi_{\dot A}$ are in the \emph{same} representation of the bosonic subalgebra of $G$; this  representation is furnished by the set of fermionic generators $Q_a$. In this
case, the $\CN=4$ supersymmetry is promoted to $\CN=5$, as first
proved in Ref. \cite{Hosomichi:2008jb}.

\section{Classification of $\CN=4$ Quiver Gauge Theories}\label{Allex}
After working out the example in section \ref{secNewEx}, it is not
difficult to find out the other gauge groups. We first review all
known examples of $\CN=4$ theories. We consider the following
pairs of superalgebras \cite{HosomichiJD, MFM:Aug09}:
\begin{eqnarray}\label{knownN4}
(G,G^\prime)&=&(U(N_1|N_2),(U(N_2|N_3)),
(OSp(N_1|2N_2),(OSp(N_1|2N_3)),
\nonumber\\&&(OSp(N_1|2N_2),(OSp(N_3|2N_2)),
(OSp(N_1|2N_2),(OSp(2|2N_2)),\nonumber\\&&
(OSp(2|2N_1),(OSp(2|2N_1)).
\end{eqnarray}
For every pair, the even parts share at least one common \emph{simple
factor}, hence can be chosen as the Lie algebras of the gauge
groups.

It is straightforward to generalize the construction of section
\ref{secN4} by letting that one symplectic 3-algebra contains
\emph{three} symplectic sub 3-algebras. And one can realize the three sub
3-algebras in terms of \emph{three} superalgebras. Then the gauge
group must be the even parts of $(G_1,G_2,G_3)$, where $G_i$
($i=1, 2,3)$ is a superalgebra selected from the list
\begin{equation}\label{listsafewer}
U(M|N),\quad OSp(M|2N),\quad  OSp(2|2N).
\end{equation}
Here we assume that the even parts of $G_1$ and $G_2$ share at least
one common simple factor, while the even parts of $G_2$ and $G_3$
share at least one common simple factor. For example, one can
choose $(G_1,G_2,G_3)$ as
$(OSp(N_1|2N_2),OSp(N_1|2N_3),OSp(N_4|2N_3))$. The resulting quiver
diagram for gauge groups is
\begin{equation}\label{linear1}
Sp(2N_2)-SO(N_1)-Sp(2N_3)-SO(N_4).
\end{equation}
Or we can set $(G_1,G_2,G_3)$= $(U(N_1|N_2),U(N_2|N_3),U(N_3|N_4))$,
and the resulting quiver diagram for gauge groups is
\begin{equation}\label{linear2}
U(N_1)-U(N_2)-U(N_3)-U(N_4).
\end{equation}
In the general case, one can choose the even parts of
$(G_1,\cdots,G_n)$, where $G_i$ ($i=1, \cdots, n)$ is a
superalgebra selected from the list (\ref{listsafewer}); the even
parts of $G_i$ and $G_{i+1}$ ($i=1,\cdots,n-1)$ share at least one
common simple factor \cite{HosomichiJD, MFM:Aug09}. If the even
parts of $G_1$ and $G_n$ (with $n$ an even number) also share at
least one common simple factor, then the linear quiver becomes a
closed loop.
The linear quiver gauge theories described in this paragraph exhaust
all known examples of $\CN=4$ superconformal CMS theories.

As \cite{MFM:Aug09} pointed out, if one also takes account of the
exceptional superalgebras, and the isomorphisms of the Lie
algebras, there are additional possibilities. We will elaborate
these ideas by constructing some $\CN=4$ theories with new gauge
groups.

Let us first consider the exceptional superalgebras. The even parts
of the superalgebras $F(4)$, $G(3)$ and $D(2|1,\alpha)$ (with $\a$ a
continuous parameter) are $SO(7)\times SU(2)$ ($SO(7)$ is in the
spinor representation), $G_2\times SU(2)$ and $SO(4)\times Sp(2)$,
respectively. Now we have the complete list
\begin{equation}\label{listsa}
U(M|N),\quad OSp(M|2N),\quad  OSp(2|2N),\quad F(4),\quad G(3),\quad
D(2|1;\a).
\end{equation}
The superalgebras $SU(M|N)$ and $PSU(2|2)$, the cousins of
$U(M|N)$, can be also used to realize the sympelctic 3-algebra. We
therefore may have
\begin{eqnarray}\label{exN4}
(G,G^\prime)&=&(F(4),SU(2|N_2)), (G(3),SU(2|N_2)), (G(3),F(4)),\\
\nonumber&& (OSp(N_1|2),D(2|1,\alpha)), (OSp(7|2N),F(4)),
(OSp(4|2N),D(2|1,\alpha)).
\end{eqnarray}
Their even parts can be selected as the Lie algebras of the gauge
groups.

It also is possible to construct some new $\CN=4$ CMS theories by
using the four isomorphisms of the Lie algebras. We know that the Lie algebra of
$SO(3)$ is isomorphic to that of $SU(2)$ and $Sp(2)$, the Lie algebra of $SO(5)$
is isomorphic to that of $Sp(4)$, and the Lie algebra of $SO(6)$
is isomorphic to that of $SU(4)$.
So the pairs of the superalgebras can be also chosen as
\begin{eqnarray}\label{isoN4}
(G,G^\prime)&=&(OSp(3|2N_1),OSp(N_2|2)), (OSp(3|2N_1),SU(2|N_2)),
(OSp(3|2N_1),F_4),\nonumber\\&& (OSp(3|2N_1),D(2|1,\alpha)),
(OSp(3|2N_1),G_3),(OSp(N_1|2),SU(2|N_2)),\nonumber\\&&
(OSp(N_1|2),F_4), (OSp(N_1|2),G_3), (G_3,D(2|1,\alpha)),
(F_4,D(2|1,\alpha)), \nonumber\\&&(OSp(5|2N_1),OSp(N_2|4)),
(OSp(6|N_1),SU(4|N_2)),(D(2|1,\alpha), SU(2|N_2)),\nonumber\\
\end{eqnarray}
and their even parts can be selected as the Lie algebras of the
gauge groups.


Finally, in Section \ref{secNewEx}, we have constructed two
classes of new $\CN=4$ theories by requiring the $U(1)$ parts of
the even parts of $G$ and $G^\prime$ are identical; they are given
by \e\label{shru1} (G,G^\prime) &=& ((OSp(2|2N_2), U(N_3|N_4)),
((U(N_1|N_2), U(N_3|N_4)). \ee The other classes of new $\CN=4$
theories in Section \ref{secNewEx} are given by the lists
(\ref{ex1}) and (\ref{ex2}).

In summary, one can use the constructions of the previous
paragraphs in the general case $(G_1,\cdots,G_n)$, where any
adjacent pair $G_i$ and $G_{i+1}$ ($i=1,\cdots,n-1)$ is selected
from (\ref{knownN4}), (\ref{exN4}), (\ref{isoN4}), (\ref{shru1}),
(\ref{ex1}), (\ref{ex2}) or (\ref{idso4}). Namely, the even parts
of $G_i$ and $G_{i+1}$ share at least one common simple factor or
$U(1)$ factor; or at least one simple factor of the even part of
$G_i$ is isomorphic to one simple factor of the even part of
$G_{i+1}$, even they may be in different representations. The Lie
algebra of the gauge group is just the even parts of the
superalgebras $(G_1,\cdots,G_n)$, and the representations are
determined by the fermionic generators. If $n$ is even, then $G_1$
and $G_n$ may also share one common bosonic parts, i.e. $G_1\sim
G_n$ may form a closed loop \cite{HosomichiJD}.

\section{Conclusions and Discussion}\label{conclusions}
In this paper, we have identified a special class of 3-algebras called double-symplectic 3-algebras used to construct the general $\CN=4$ quiver gauge theories, and showed that its
consistent contraction gives a class of 3-algebras called $\CN=4$ three-algebras, which can
be also used to construct the general $\CN=4$ theories.

We then have used two superalgebras whose bosonic parts
share at least one simple factor or $U(1)$ factor to construct the four sets of 3-brackets (\ref{mbarcks0}) and (\ref{mbrcks00}) and the four sets of
FIs (\ref{FI4}) defined on the
two sub symplectic 3-algebras in the $\CN=4$ theories. We have also generalized the
construction to the more general $\CN=4$ quiver gauge theories in which
more than two sub 3-algebras are used. We have not only rederived
all known $\CN=4$ theories, but also constructed many classes
of $\CN=4$ quiver gauge theories (see Sec. \ref{secNewEx} and Sec.
\ref{Allex}) with new gauge groups. Especially, if the common parts of the bosonic parts
of two superalgebras are $U(1)$ factors, the resulting $\CN=4$
theories can contain free parameters. We have constructed two
infinite classes of new theories of this kind (see Sec.
\ref{sec4un} and Sec. \ref{secospu}). It would be interesting to
see whether or not there are further constraints imposed on these
free parameters if we quantize these theories.

Taking account of the new
$\CN=4$ theories, we have been able to achieve a new classification of all possible $\CN=4$ quiver gauge theories in Sec. \ref{secospu}.

Using superalgebras to realize the 3-algegbra, we have showed the general $\CN=4$
 theory in terms of the double-symplectic 3-algebra is equivalent to the previous $\CN=4$ theory \cite{HosomichiJD}
 derived directly from the conventional Lie 2-algebra.
Hence the $\CN=4$ theories with new gauge groups in Sec.
\ref{secNewEx} derived in the 3-algebra framework can be
also understood as special examples of the general $\CN=4$ HLLLP
theories in Ref. \cite{HosomichiJD}. Specifically, in \cite{HosomichiJD},
one can choose either $SO(4)\cong SU(2)\times SU(2)$ or $U(N)\cong SU(N)\times U(1)$
as the common bosonic algebra of the two superalgebras; our constructions in
Sec. \ref{secNewEx} show that one can also choose only $SU(2)$ (the indecomposable
part of $SO(4)$) or $U(1)$ (the indecomposable  part of $U(N)$) as the common bosonic part of the two superalgebras.

Most of gravity duals of these $N=4$ quiver gauge theories have
not been constructed yet. It would be nice to study the quantum
properties of these $N=4$ quiver gauge theories and to construct
and study the corresponding gravity duals.

Finally, we know that it was demonstrated that generic
Chern-Simons gauge theories with or without (massless) matter are
conformally invariant even at the quantum level \cite{CSW1, CSW0,
CSW2, Piguet, Saemann1}. It would be interesting to examine the
conformal invariance of these $\CN=4$ quiver gauge theories at
quantum levels.

\section{Acknowledgement}

We thank the referee for comments. YSW is supported in part by the US NSF through Grant No.
PHY-0756958 and PHY-1068558. FMC is supported by the China Postdoctoral Science Foundation through Grant No. 2012M510244.

\appendix

\section{Conventions and Useful Identities}\label{Identities}
The conventions and useful identities are adopted from our previous
paper \cite{ChenWu3}.

\subsection{Spinor Algebra}
In $1+2$ dimensions, the gamma matrices are defined as
\begin{equation}
(\gamma_{\mu})_{\alpha}{}^\gamma(\gamma_{\nu})_{\gamma}{}^\beta+
(\gamma_{\nu})_{\alpha}{}^\gamma(\gamma_{\mu})_{\gamma}{}^\beta=
2\eta_{\mu\nu}\delta_{\alpha}{}^\beta.
\end{equation} For the metric we
use the $(-,+,+)$ convention. The gamma matrices in the Majorana
representation can be defined in terms of Pauli matrices:
$(\gamma_{\mu})_{\alpha}{}^\beta=(i\sigma_2, \sigma_1, \sigma_3)$,
satisfying the important identity
\begin{equation}
(\gamma_{\mu})_{\alpha}{}^\gamma(\gamma_{\nu})_{\gamma}{}^\beta
=\eta_{\mu\nu}\delta_{\alpha}{}^\beta+\varepsilon_{\mu\nu\lambda}(\gamma^{\lambda})_{\alpha}{}^\beta.
\end{equation}
We also define
$\varepsilon^{\mu\nu\lambda}=-\varepsilon_{\mu\nu\lambda}$. So
$\varepsilon_{\mu\nu\lambda}\varepsilon^{\rho\nu\lambda} =
-2\delta_\mu{}^\rho$. We raise and lower spinor indices with an
antisymmetric matrix
$\epsilon_{\alpha\beta}=-\epsilon^{\alpha\beta}$, with
$\epsilon_{12}=-1$. For example,
$\psi^\alpha=\epsilon^{\alpha\beta}\psi_\beta$ and
$\gamma^\mu_{\alpha\beta}=\epsilon_{\beta\gamma}(\gamma^\mu)_\alpha{}^\gamma
$, where $\psi_\beta$ is a Majorana spinor. Notice that
$\gamma^\mu_{\alpha\beta}=(\mathbbm{l}, -\sigma^3, \sigma^1)$ are
symmetric in $\alpha\beta$. A vector can be represented by a
symmetric bispinor and vice versa:
\begin{equation}
A_{\alpha\beta}=A_\mu\gamma^\mu_{\alpha\beta},\quad\quad A_\mu=-\frac{1}{2}\gamma^{\alpha\beta}_\mu A_{\alpha\beta}.
\end{equation}
We use the following spinor summation convention:
\begin{equation}
\psi\chi=\psi^\alpha\chi_\alpha,\quad\quad
\psi\gamma_\mu\chi=\psi^\alpha(\gamma_{\mu})_{\alpha}{}^\beta\chi_\beta,
\end{equation}
where $\psi$ and $\chi$ are anti-commuting Majorana spinors. In
$1+2$ dimensions the Fierz transformation reads
\begin{eqnarray}
(\lambda\chi)\psi &=& -\frac{1}{2}(\lambda\psi)\chi -\frac{1}{2}
(\lambda\gamma_\nu\psi)\gamma^\nu\chi.
\end{eqnarray}

\subsection{$SU(2)\times SU(2)$ Identities}\label{SO4}
We define the 4 sigma matrices as
\begin{equation}\label{pulim}
\sigma^a{}_A{}^{\dot{B}}=(\sigma^1,\sigma^2,\sigma^3,i\mathbbm{l}),
\end{equation}
by which one can establish a connection between the $SU(2)\times
SU(2)$ and $SO(4)$ group. These sigma matrices satisfy the following
Clifford algebra:
\begin{eqnarray}
\sigma^a{}_{A}{}^{\dot{C}}\sigma^{b\dag}{}_{\dot{C}}{}^B+
\sigma^b{}_{A}{}^{\dot{C}}\sigma^{a\dag}{}_{\dot{C}}{}^B=2\delta^{ab}\delta_A{}^B,\\
\sigma^{a\dag}{}_{\dot{A}}{}^{C}\sigma^{b}{}_{C}{}^{\dot{B}}+
\sigma^{b\dag}{}_{\dot{A}}{}^{C}\sigma^{a}{}_{C}{}^{\dot{B}}=2\delta^{ab}\delta_{\dot{A}}{}^{\dot{B}}.
\end{eqnarray}
We use anti-symmetric matrices
\begin{eqnarray}
\epsilon_{AB}=-\epsilon^{AB}=\begin{pmatrix} 0&-1 \\ 1&0
\end{pmatrix}\;
\;\;{\rm and}\;\;\;
\epsilon_{\dot{A}\dot{B}}=-\epsilon^{\dot{A}\dot{B}}=\begin{pmatrix}
0&1 \\-1& 0
\end{pmatrix}
\end{eqnarray}
to raise or lower un-dotted and dotted indices, respectively. For
example,
$\sigma^{a\dag\dot{A}B}=\epsilon^{\dot{A}\dot{B}}\sigma^{a\dag}{}_{\dot{B}}{}^{B}$
and $\sigma^{aB\dot{A}}=\epsilon^{BC}\sigma^{a}{}_{C}{}^{\dot{A}}$.
The sigma matrix $\sigma^a$ satisfies a reality condition
\begin{equation}\label{RC4}
\sigma^{a\dag}{}_{\dot{A}}{}^{B}=-\epsilon^{BC}\epsilon_{\dot{A}\dot{B}}\sigma^a{}_{C}{}^{\dot{B}},\quad
{\rm or} \quad\sigma^{a\dag\dot{A}B}=-\sigma^{aB\dot{A}}.
\end{equation}
The antisymmetric matrix $\epsilon_{AB}$ satisfies an important
identity
\begin{equation}
\epsilon_{AB}\epsilon^{CD}=-(\delta_A{}^C\delta_B{}^{D}-\delta_A{}^D\delta_B{}^{C}),
\end{equation}
and $\epsilon_{\dot{A}\dot{B}}$ satisfies a similar identity.

The parameter for the $\CN=4$ supersymmetry transformations is
defined as $\epsilon^{A\dot{B}}=\ep_a\s^{aA\dot{B}}$.

\section{The Commutation Relations of Superalgebras}\label{superalgebras}
\subsection{$U(M|N)$}\label{cmrunum}
The commutation relations of $U(M|N)$ are given by
\e\label{unum}\nonumber
&&[M_{\bu}{}^{\bv},M_{\bw}{}^{\bt}]=\d_{\bw}{}^{\bv}M_{\bu}{}^{\bt}-\d_{\bu}{}^{\bt}M_{\bw}{}^{\bv},\quad [M_\pri{}^{\pj},M_\pk{}^\pl]=\d_\pk{}^\pj M_\pri{}^\pl-\d_\pri{}^\pl M_{\pk}{}^\pj\nonumber\\
&&[M_{\bu}{}^{\bv},Q_{\bw}{}^\pk]=\d_{\bw}{}^{\bv}Q_{\bu}{}^\pk,\quad [M_{\bu}{}^{\bv},\bar Q_\pk{}^{\bw}]=-\d_{\bu}{}^{\bw}\bar Q_\pk{}^{\bv},\nonumber\\
&&[M_\pri{}^{\pj},Q_{\bw}{}^\pk]=-\d_\pri{}^\pk Q_{\bw}{}^\pj,\quad [M_\pri{}^{\pj},\bar Q_\pk{}^{\bw}]=\d_\pk{}^\pj \bar Q_\pri{}^{\bw}\nonumber\\
&& \{Q_{\bu}{}^\pri,\bar Q_\pj{}^{\bv}\}=k(\d_\pj{}^\pri M_{\bu}{}^{\bv}+\d_{\bu}{}^{\bv}M_{\pj}{}^\pri),
\ee
where $Q_{\bu}{}^\pri$ carries a $U(M)$ fundamental index $\bu=1,\cdots,M$ and a $U(N)$ anti-fundamental index $\pri=1,\cdots,N$.
Here we have
\e\label{dcmpq}
Q_\pa=\begin{pmatrix} \bar Q_\pri{}^{\bu}\\-Q_{\bu}{}^\pri\end{pmatrix}=\bar Q_\pri{}^{\bu}\d_{1\a}-Q_{\bu}{}^\pri\d_{2\a},
\ee
 In the second equation of (\ref{dcmpq}), we have introduced a ``spin up" spinor $\chi_{1\alpha}$ and a ``spin
down" spinor $\chi_{2\alpha}$, i.e., \footnote{Here the index $\a$
is \emph{not} a spacetime spinor index. We hope this will not
cause any confusion.}
\begin{eqnarray}
\chi_{1\alpha}=\begin{pmatrix} 1 \\ 0
\end{pmatrix}=\d_{1\a}\;
\;\;{\rm and}\;\;\; \chi_{2\alpha}=\begin{pmatrix} 0 \\ 1
\end{pmatrix}=\d_{2\a}.
\end{eqnarray}
And the anti-symmetric tensor $\om_{ab}$ and its inverse read
\begin{equation}\label{dcmo}
\omega_{\pa\pb}=\begin{pmatrix} 0 & \d_{\bv}{}^{\bu}\d_{\pri}{}^\pj \\
-\d^{\bv}{}_{\bu}\d^{\pri}{}_\pj & 0
\end{pmatrix},\quad \omega^{\pb\pc}=\begin{pmatrix} 0 & -\d^{\pj}{}_\pk\d^{\bw}{}_{\bv} \\
\d_{\pj}{}^\pk\d_{\bw}{}^{\bv} & 0
\end{pmatrix}.
\end{equation}
With (\ref{dcmpq}) and (\ref{dcmo}), the superalgebra (\ref{unum}) takes the form of (\ref{slie2}) or (\ref{slie3}).

\subsection{$OSp(M|2N)$}\label{cmrsp22n}
The superalgebra $OSp(M|2N)$ reads
\begin{eqnarray}\label{OSp}
&&[M_{\bar i\bar j},M_{\bar k\bar
l}]=\d_{\bj\bk}M_{\bi\bl}-\d_{\bi\bk}M_{\bj\bl}+\d_{\bi\bl}M_{\bj\bk}
-\d_{\bj\bl}M_{\bi\bk},\nonumber\\
&&[M_{\hi\hj},M_{\hk\hl}]=\omega_{\hj\hk}M_{\hi\hl}+\omega_{\hi\hk}M_{\hj\hl}
+\omega_{\hi\hl}M_{\hj\hk}+\omega_{\hj\hl}M_{\hi\hk},\nonumber\\
&&[M_{\bi\bj},Q_{\bk\hk}]=\d_{\bj\bk}Q_{\bi\hk}-\d_{\bi\bk}Q_{\bj\hk},\nonumber\\
&&[M_{\hi\hj},Q_{\bk\hk}]=\omega_{\hj\hk}Q_{\bk\hi}+\omega_{\hi\hk}Q_{\bk\hj},\nonumber\\
&&\{Q_{\bi\hi},Q_{\bj\hj}\}=k(\omega_{\hi\hj}M_{\bi\bj}+\d_{\bi\bj}M_{\hi\hj}),
\end{eqnarray}
where $\bi=1,\cdots,M$ is an $SO(M)$ fundamental index, and
$\hi=1,\cdots, 2N$ an $Sp(2N)$ fundamental index. Here we have
\begin{equation}
Q_a=Q_{\bi\hi} \quad {\rm and}\quad
\omega_{ab}=\omega_{\bi\hi,\bj\hj}=\d_{\bi\bj}\omega_{\hi\hj}.
\end{equation}
Now the superalgebra (\ref{cmrsp22n}) also takes the form of (\ref{slie2}) or (\ref{slie3}).

\end{document}